\documentclass{aastex62}

\usepackage[%
        ]{hyperref}

\hypersetup{
	hypertexnames=true,
	plainpages=false,
	naturalnames=true}
\usepackage{textcomp}
\usepackage{savesym}
\savesymbol{tablenum}
\usepackage{siunitx}
\restoresymbol{SIX}{tablenum}
\usepackage{gensymb}
\usepackage{comment}
\usepackage[utf8]{inputenc}
\usepackage{array}
\usepackage{float}
\usepackage{graphicx}
\usepackage{footmisc}
\usepackage{multirow}
\usepackage{subfigure}
\shorttitle{} 
\shortauthors{Suissa et al.}

\begin{document}
\label{firstpage}

\title{Dim Prospects for Transmission Spectra of Ocean Earths Around M Stars}
\accepted{to ApJ}

\correspondingauthor{Gabrielle Suissa}
\email{gabrielle.engelmann-suissa@nasa.gov}

\author[0000-0003-4471-1042]{Gabrielle Suissa}
\affiliation{Dept. of Astronomy, Columbia University, 550 W 120th Street, New York NY 10027}
\affiliation{NASA Goddard Space Flight Center, 8800 Greenbelt Rd, Mail Stop 699.0 Building 34, Greenbelt, MD 20771, USA}
\affiliation{NASA Goddard Sellers Exoplanet Environments Collaboration}
\affiliation{Goddard Earth Sciences Technology and Research (GESTAR), Universities Space Research Association, Columbia, Maryland, USA}

\author[0000-0002-8119-3355]{Avi M. Mandell}
\affiliation{NASA Goddard Space Flight Center, 8800 Greenbelt Rd, Mail Stop 693.0 Building 34, Greenbelt, MD 20771, USA}
\affiliation{NASA Goddard Sellers Exoplanet Environments Collaboration}

\author[0000-0002-7188-1648]{Eric T. Wolf}
\affiliation{Laboratory for Atmospheric and Space Physics, Department of Atmospheric and Oceanic Sciences, University of Colorado Boulder, Boulder, CO 80309, USA}
\affiliation{NASA Goddard Sellers Exoplanet Environments Collaboration}
\affiliation{NASA NExSS Virtual Planetary Laboratory, Box 951580, Seattle, WA 98195}

\author[0000-0002-2662-5776]{Geronimo L. Villanueva}
\affiliation{NASA Goddard Space Flight Center, 8800 Greenbelt Rd, Mail Stop 693.0 Building 34, Greenbelt, MD 20771, USA}
\affiliation{NASA Goddard Sellers Exoplanet Environments Collaboration}

\author[0000-0002-5967-9631]{Thomas Fauchez}
\affiliation{NASA Goddard Space Flight Center, 8800 Greenbelt Rd, Mail Stop 699.0 Building 34, Greenbelt, MD 20771, USA}
\affiliation{Goddard Earth Sciences Technology and Research (GESTAR), Universities Space Research Association, Columbia, Maryland, USA}
\affiliation{NASA Goddard Sellers Exoplanet Environments Collaboration}

\author[0000-0002-5893-2471]{Ravi kumar Kopparapu}
\affiliation{NASA Goddard Space Flight Center, 8800 Greenbelt Rd, Mail Stop 699.0 Building 34, Greenbelt, MD 20771, USA}
\affiliation{NASA Goddard Sellers Exoplanet Environments Collaboration}
\affiliation{NASA NExSS Virtual Planetary Laboratory, Box 951580, Seattle, WA 98195}

\begin{abstract}
The search for water-rich Earth-sized exoplanets around low-mass stars is rapidly gaining attention because they represent the best opportunity to characterize habitable planets in the near future. Understanding the atmospheres of these planets and determining the optimal strategy for characterizing them through transmission spectroscopy with our upcoming instrumentation is essential in order to constrain their environments. For this study, we present simulated transmission spectra of tidally locked Earth-sized ocean-covered planets around late-M to mid-K stellar spectral types, utilizing GCM modeling results previously published by \cite{ravi:2017} as inputs for our radiative transfer calculations performed using NASA’s Planetary Spectrum Generator (\url{psg.gsfc.nasa.gov}; \cite{psg:2018}). We identify trends in the depth of H$_2$O spectral features as a function of planet surface temperature and rotation rate. These trends allow us to calculate the exposure times necessary to detect water vapor in the atmospheres of aquaplanets through transmission spectroscopy with the upcoming James Webb Space Telescope (JWST) as well as several future flagship space telescope concepts under consideration (LUVOIR and OST) for a target list constructed from the TESS Input Catalog (TIC). Our calculations reveal that transmission spectra for water-rich Earth-sized planets around low-mass stars will be dominated by clouds, with spectral features $<$ 20~ppm, and only a small subset of TIC stars would allow for the characterization of an ocean planet in the Habitable Zone. We thus present a careful prioritization of targets that are most amenable to follow-up characterizations with next-generation instrumentation, in order to assist the community in efficiently utilizing precious telescope time. 

\end{abstract}


\section{Introduction}
\label{sec:introduction}
The past decade has seen extraordinary progress in our understanding of extrasolar planets, including their occurrence rates, orbital dynamics, and density constraints. Thanks largely to both ground and spaced-based surveys, including the Kepler and TESS (The Transiting Exoplanet Survey Satellite) missions, we are now aware of over 4000 planets beyond our solar system\footnote{\url{https://exoplanetarchive.ipac.caltech.edu/}}, and we expect to discover thousands more in the near future \citep{sullivan:2015,barclay:2018}. The new frontier in the exoplanetary science will soon transition to the characterization of the atmospheres of the best transiting exoplanet targets. The James Webb Space Telescope (JWST), slated to launch in 2021, will be the primary platform for transit spectroscopy in the coming decade, with support from ground-based telescopes. Potential next-generation flagship missions currently being proposed, such as LUVOIR (the Large UV Optical Infrared Surveyor) and OST (the Origins Space Telescope), would have improved sensitivity and similar or greater collecting area compared with JWST. 

Some of the most exciting candidates that are amenable for atmospheric characterization are Earth-sized planets orbiting cool stars (late-K and M dwarfs) in the Habitable Zone (HZ, i.e. the surface-water zone). These systems' planet-to-star radius ratios are relatively high and can therefore yield deeper spectral features in transmission spectroscopy. For terrestrial planets orbiting these low-mass stars in the HZ, their frequent transits can also make characterization more feasible with limited observatory lifetimes, compared to HZ planets around higher-mass stars. Furthermore, obtaining a spectrum of a planet residing within the HZ that can provide confirmation of an ocean-covered surface and allow for the detection of biosignatures would have incredible implications for the search for habitable worlds.

In this work, we take into account that the atmospheres of terrestrial planets in the HZ of late-M to mid-K dwarfs cannot be assumed to be true Earth twins because they are expected to be tidally locked \citep{dobrovolskis:2009,barnes:2013,barnes:2017}. If the planet's orbital eccentricity is near zero, this can result in synchronous rotation \citep{leconte:2015}, in which one side of the planet always faces the star. Recent 3-D climate modeling of slow and synchronously rotating Earth-sized planets with oceans around M stars \citep{yang:2013,yang:2014,hu:2014,way:2015,way:2016,ravi:2016,ravi:2017,wolf:2017,fujii:2017,haqq:2018,bin:2018,chen:2018,komacek:2019,wolf:2019} has shown that the atmospheric circulation changes significantly due to their slower rotation than that of the Earth. Specifically, slow rotation weakens the Coriolis effect and causes the atmospheric circulation to shift from a ``rapidly rotating" regime to a ``slowly rotating" regime --- that is, from an atmosphere with a significant zonal component to the circulation and banded cloud formations symmetric around the equator, to an atmosphere where heat is transported radially from the day side to the night side and circular clouds develop symmetrically about the substellar point. These thick clouds at the substellar point in the slowly rotating regime cool the planet by significantly increasing the planetary albedo, thus making the planet maintain habitable conditions (i.e. low surface temperatures) at higher incident stellar fluxes than is possible otherwise. Another notable consequence of the change in circulation regimes is that for slow rotators, water vapor is lofted higher into the atmosphere by strong up-welling motions at the substellar point, resulting in significantly higher stratospheric water vapor mixing ratios compared to fast rotating planets (assumed identical surface temperatures). The general circulation model (GCM) simulations conducted by \citet{ravi:2017} represent aquaplanets around late-M to mid-K stars with a wide range of incident stellar fluxes and rotation rates. They self-consistently capture the complex interplay between stellar irradiance, atmospheric circulation, water vapor, clouds, and temperature (see Section~\ref{sec:gcm}). 

As the TESS mission discovers more Earth-sized planets, the exoplanet community will need a cohesive strategy to decide which planets should be priorities for atmospheric characterization. Although we have promising instrumentation for our flagship missions, we are limited by constraints on telescope time and our current technological efficiencies. One of the challenges will be anticipating if ocean-covered Earth-like planets are amenable to our current capabilities for identifying planetary atmospheres, and if so, which candidates are the most suitable for our telescopes to discover such a planet. Our goal in this paper is to thus qualify how successful JWST, and the proposed LUVOIR (Architectures A and B) and Origins Space Telescope missions, will be in identifying the atmospheric properties of these synchronously rotating Earth-like planets orbiting cool stars that TESS is expected to discover in the coming years. Based on the GCM results from \cite{ravi:2017} we predict what a spectrum of such an aquaplanet would look like, in order to calculate how feasible (via exposure time) it would be for our upcoming/proposed space telescopes to detect it and identify it as an ocean Earth. We note that our exposure times are only relevant if the planet is synchronously rotating around the star, as it is a possibility that HZ planets orbiting K stars are in other rotations states, such as spin-orbit resonance \citep{leconte:2015}. We also present a priority list of the dwarf stars that, given they are found to have a terrestrial planet orbiting them in the HZ, would be the best candidates for this characterization.

The structure of this paper is as follows. In Section~\ref{sec:methods}, we detail the GCM models and radiative transfer tool used. Section~\ref{sec:observatories} highlights the water vapor features found in the resulting transmission spectrum and correlates them to the spectral capabilities of JWST, LUVOIR, and OST. Section~\ref{sec:tess} outlines our criteria in selecting stellar targets from the TESS Input Catalog. We present and analyze model transmission spectra for a range of terrestrial-planets/cool-star systems in Sections~\ref{sec:analysis} and \ref{sec:trends}. In Sections~\ref{sec:tess} and \ref{sec:calc-exptimes}, we calculate exposure times for ocean Earths given they orbit a cool star being currently observed by TESS. We compare the atmospheric characterization capabilities of different telescopes and discuss the implications of this work in Section~\ref{sec:discussion}.

\section{Methods}
\label{sec:methods}

\subsection{GCM Model}
\label{sec:gcm}
The general circulation model (GCM) outputs that we use for this work were simulated by \cite{ravi:2017}. Briefly, \cite{ravi:2017} use a modified version of the Community Atmosphere Model (CAM) version 4 \citep{CAM}, called ExoCAM\footnote{\url{https://github.com/storyofthewolf/ExoCAM}}$^,$\footnote{\url{https://github.com/storyofthewolf/ExoRT}} that is suitable for exoplanet habitability studies. More details about the updates made to ExoCAM are given in \citet{wolf:2017,haqq:2018,wolf:2019}. \cite{ravi:2017} explored 39 different configurations of an ocean-covered Earth-sized planet synchronously rotating around a late-M to mid-K star ($\mathrm{T_{eff}}$ from 2600~K to 4500~K) in the HZ. The simulated planets experience a range of varying incident fluxes from their host stars. The effective temperature, luminosity, period, and mass of the star are defined a priori, while the stellar radius is calculated using the Stefan–Boltzmann Law $( L = 4\pi \sigma R^2 T^4)$ for consistency. The planets in every simulation share a radius and mass of 1R$_\oplus$ and 1M$_\oplus$. They are fixed to be entirely covered with a 50~m slab of ocean without any ocean heat transport. Their atmospheres assume a surface pressure of 1~bar, like the Earth, and include just H$_2$O and N$_2$ (CO$_2$ is removed for simplicity as the original study was focused on water vapor and cloud processes near the inner edge of HZ; see Section~\ref{sec:spectral features}). Both liquid and ice water clouds are prognostically calculated in the model based on the local atmospheric conditions in each grid-cell. Clouds are treated as Mie-scatterers with liquid cloud droplet radii assumed to be 14~$\mathrm{\micro}$m in all atmospheric layers. Ice clouds have ice crystal effective radii ranging from a few tenths to a few hundred microns based on a temperature-dependent parameterization. 

3-D climate simulations were run until either they reached thermal equilibrium (i.e. the net absorbed stellar flux equals the outgoing longwave flux and the global mean temperature has stabilized) or until a runaway greenhouse was triggered via a critical energy imbalance due to overwhelming thermal water vapor opacities \citep{goldblatt:2013}. The last converged solutions (i.e. climatologically stable) were taken as the inner edge of the HZ \citep{ravi:2017}. Note that a fully realized runaway greenhouse would evaporate the entirety of the oceans into its atmospheres leading to surface temperatures of $\sim$1600~K \citep{goldblatt:2013}. However, such a climate is beyond the operational bounds of the GCM, which becomes numerically unstable when global mean surface temperatures approach $\sim$400~K. For these cases, \citet{ravi:2017} took the 3-D model results from the last several orbits before the numerical instability occurred. Thus, the “runaway” cases represent a snapshot of planets that are undergoing a runaway greenhouse process, and are therefore in a transient state. With respect to observable phenomena, these incipient runaway states are perhaps best considered as a sampling of hot, moist, and optically thick atmospheres (e.g., \cite{goldblatt:2015}). 

The \cite{ravi:2017} results also indicate that some planets reach moist greenhouse states ($>$10$^{-3}$ H$_2$O mixing ratio at 1~mbar) with relatively habitable surface temperatures ($\sim$280~K) due to enhanced stratospheric water vapor driven by the general circulation of slow rotators. Note that \citet{fujii:2017} arrived at a similar result, arguing that increased near-IR absorption also contributed to elevated stratospheric water vapor mixing ratios. A list of the simulations implemented by \citet{ravi:2017} and their stellar-planetary characteristics can be found in Table~\ref{tab:sims}. 

We do not model the effects of stellar UV activity and the corresponding photochemistry in this work. Recently, \citet{chen:2018} and \citet{afrin:2019} discussed the effects of changes in stellar activity on the atmospheric chemistry of HZ planets around M dwarfs in the context of biosignatures. In particular, \citet{afrin:2019} use the same GCM simulations as we do (from \citet{ravi:2017}) to assess the impact of varying UV activity on the detectability of water-vapor features in the transmission spectrum. We will discuss our results in the context of \citet{afrin:2019} in Section~\ref{sec:discussion}. 

The model top for the GCMs conducted by \citet{ravi:2017} lies at 1~mbar. This model top reaches up to about Earth's stratopause, and captures the important processes for regulation of a planet's large-scale energy balance and thus its climate. However, for simulating transmission spectra (see Section~\ref{sec:psg}), a greater vertical extent is often needed in order to fully capture the radiative interactions that occur during a transit event. For terrestrial climates of approximately Earth temperatures and below, a 1~mbar model top still can yield adequate results for transmission spectra. However, for warm water-rich planets ($>$300~K) a 1~mbar model becomes increasingly insufficient for properly capturing transmission spectra because the opacity of the atmosphere increases and pushes transit depths to lower atmospheric pressures. Therefore, before calculating transmission spectra, the GCM model columns must be artificially extended to as high as 1~$\micro$bar. Otherwise, in these hot climates the cloud layers can be pushed beyond the GCM model top, and their information can become skewed or lost entirely. Information about clouds that may exist in these low-pressure regimes, however, is essential for understanding the warm atmospheres of planets undergoing a runaway state and their transmission spectra. We explore several different assumptions for approximating the clouds in the extended layers of the atmospheric profiles, detailed in Section~\ref{sec:modeltop}. We demonstrate that for hot aquaplanet spectroscopy it is crucial to understand cloud processes in these lower-pressure regimes. 

\subsection{The Planetary Spectrum Generator}
\label{sec:psg}

In order to create synthesized spectra from our GCM outputs, we use the Planetary Spectrum Generator (PSG), a radiative transfer tool publicly available online at \url{https://psg.gsfc.nasa.gov/} \citep{psg:2018}. PSG combines line-by-line modeling with a robust scattering model for aerosols in order to both efficiently and precisely calculate planetary spectra. Drawing upon spectral line data repositories for various molecular species, multiple scattering models, and the latest radiative transfer methods, PSG can yield radiance values across a broad range of wavelengths. PSG also includes a realistic noise simulator, discussed in Section~\ref{sec:calc-exptimes}. To create each spectrum, we input basic parameters of the star and planet, along with vertical profiles of the planetary atmosphere. These profiles include temperature, pressure, altitude, mass mixing ratios and particle sizes for liquid and ice clouds, and volume mixing ratios for H$_2$O and N$_2$ gas -- all outputs from the original GCM model combined with our upper-layer approximations (see Section~\ref{sec:modeltop}). PSG treats the ice clouds as complex scatterers, and the optical properties for both the ice and liquid water clouds are derived from the HITRAN~2016 database \citep{massie:2013}. In order to produce a realistic representation of the planet's atmosphere as viewed through transmission spectroscopy, we use PSG to calculate a spectrum at every interval of 4$\degree$ in latitude across the terminator of the planet (the terminator is the only part of the planet that is accessible through transmission spectroscopy). We then average the spectra over all latitude intervals, with equal weighting for each interval, to produce the final spectrum for the planet. 

\begin{table}
\centering
\caption{Parameters for the different simulated cases run by \citet{ravi:2017}. Stellar radius is included as calculated using the Stefan–Boltzmann Law, along with the global surface temperature of the planet. The final column denotes the amount of water vapor available at the original model top, and also marks the incipient runaway cases.}
\label{tab:sims}
\setlength{\textheight}{5pt}
\begin{tabular}{c c c c c c}
\hline
\hline
 $\mathrm{T_{eff}}$ [K] & Stellar Radius [R$_\odot$] & Incident Flux [W/m$^2$] & Period [days] & Surface Temperature [K] & H$_2$O Mixing Ratio [at 1~mbar] \\
\hline
2600 & 0.110 & 1200 & 4.51 & 264.75 & $7.71 \times 10^{-7}$ \\
 & & 1250 & 4.37 & 275.71 & $3.13 \times 10^{-6}$ \\
 & & 1300 & 4.25 & 284.97 & $1.18 \times 10^{-5}$ \\ 
 & & 1350 & 4.13 & 300.86 & $6.94 \times 10^{-5}$ \\
 & & 1375 & 4.07 & 342.45 & [runaway] \\ 
 & & 1400 & 4.02 & 361.82 & [runaway] \\
\hline
3000 & 0.158 & 1300 & 8.83 & 257.96 & $4.14 \times 10^{-5}$ \\ 
 & & 1400 & 8.35 & 265.97 & $2.12 \times 10^{-4}$ \\ 
 & & 1500 & 7.93 & 270.23 & $2.57 \times 10^{-4}$ \\ 
 & & 1550 & 7.74 & 275.59 & $5.14 \times 10^{-4}$ \\
 & & 1575 & 7.65 & 280.47 & $6.39 \times 10^{-4}$ \\
 & & 1600 & 7.56 & 377.81 & [runaway] \\ 
\hline
3300 & 0.302 & 1400 & 22.13 & 254.39 & $3.41 \times 10^{-5}$ \\ 
 & & 1600 & 20.02 & 266.71 & $1.24 \times 10^{-4}$ \\ 
 & & 1650 & 19.57 & 275.67 & $5.55 \times 10^{-4}$ \\ 
 & & 1700 & 19.13 & 293.98 & $7.60 \times 10^{-3}$ [water loss]\\ 
 & & 1750 & 18.72 & 362.18 & [runaway] \\ 
 & & 1800 & 18.33 & 324.41 & [runaway] \\ 
\hline
3700 & 0.505 & 1500 & 44.33 & 253.25 & $2.19 \times 10^{-5}$ \\
 & & 1600 & 42.24 & 259.25 & $7.85 \times 10^{-5}$ \\ 
 & & 1800 & 38.66 & 282.10 & $2.89 \times 10^{-3}$ [water loss] \\ 
 & & 1900 & 37.13 & 308.93 & $2.82 \times 10^{-2}$ \\ 
 & & 1950 & 36.41 & 333.18 & [runaway] \\ 
 & & 2000 & 35.73 & 353.44 & [runaway] \\
\hline
4000 & 0.618 & 1600 & 65.79 & 253.13 & $1.76 \times 10^{-5}$ \\ 
 & & 1800 & 60.23 & 266.87 & $3.61 \times 10^{-4}$ \\ 
 & & 1900 & 57.83 & 279.24 & $1.48 \times 10^{-3}$ \\ 
 & & 2000 & 55.65 & 301.08 & $1.64 \times 10^{-2}$ [water loss] \\ 
 & & 2050 & 54.63 & 329.98 & [runaway] \\ 
 & & 2100 & 53.65 & 330.68 & [runaway] \\
 & & 2200 & 51.81 & 348.21 & [runaway] \\
\hline
4500 & 0.716 & 1800 & 100.20 & 253.21 & $2.97 \times 10^{-6}$ \\
 & & 2000 & 92.59 & 262.92 & $7.04 \times 10^{-5}$ \\ 
 & & 2200 & 86.20 & 281.23 & $1.41 \times 10^{-3}$ [water loss] \\ 
 & & 2250 & 84.76 & 291.63 & $7.41 \times 10^{-3}$ \\ 
 & & 2300 & 83.37 & 303.64 & $1.79 \times 10^{-2}$ \\ 
 & & 2350 & 82.04 & 337.93 & [runaway] \\
 & & 2400 & 80.75 & 352.77 & [runaway] \\ 
 & & 2500 & 78.32 & 351.58 & [runaway] \\ 
\hline
\hline
\end{tabular}
\end{table}

\subsection{Model Top Treatment}
\label{sec:modeltop}
\begin{figure}
     \centering
     \caption{Profiles for both a non-runaway and an incipient runaway case displaying all three methods used to extend the model top.}
     \subfigure[Vertical profiles for an example of a simulation that does not enter an incipient runaway state. Clouds are localized to a narrow pressure range at altitudes where T $< 270$~K (liquid) and P $<10^{-1.5}$~bars (ice). We extend the model top from $10^{-3}$ up to $10^{-6}$~bars with three different assumptions for clouds (zero, iso and intermediate), in order to accurately capture absorption from the upper atmosphere (see Section~\ref{sec:modeltop}). For non-runaway cases the three assumptions produce the same result since the abundances for both liquid and ice clouds drop to essentially zero.]{\includegraphics[width=0.8\textwidth]{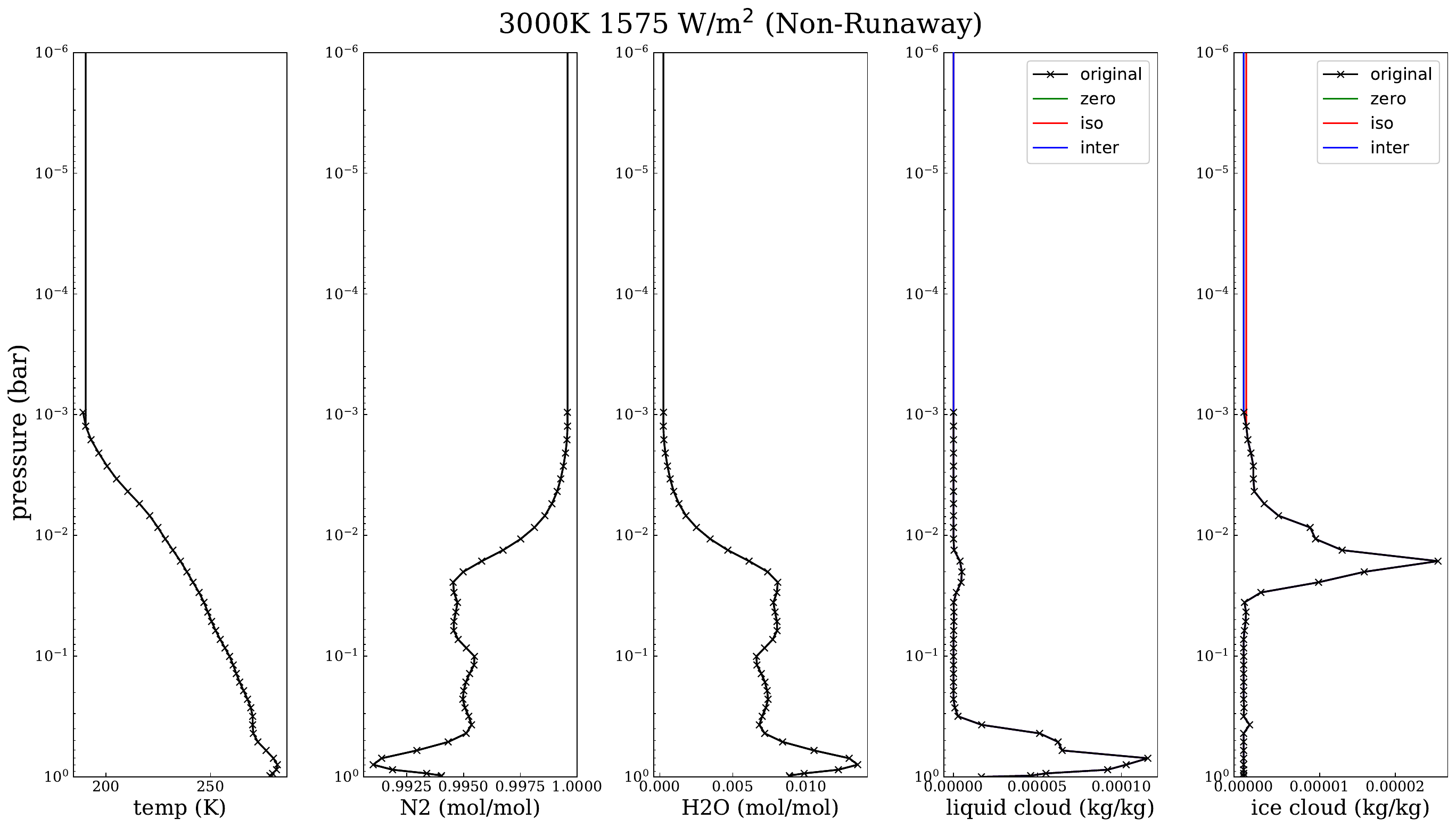}}\label{fig:nonrunaway-profiles}
     
     \subfigure[Vertical profiles for an example of a simulation that enters an incipient runaway regime. The temperature only reaches 270~K above $\sim10^{-2}$~bars, and therefore both the liquid and ice cloud profiles only provide a lower-cloud pressure below the GCM model top. The three different assumptions for extending the model top therefore produce very different cloud profiles above $10^{-3}$~bars.]{\includegraphics[width=0.8\textwidth]{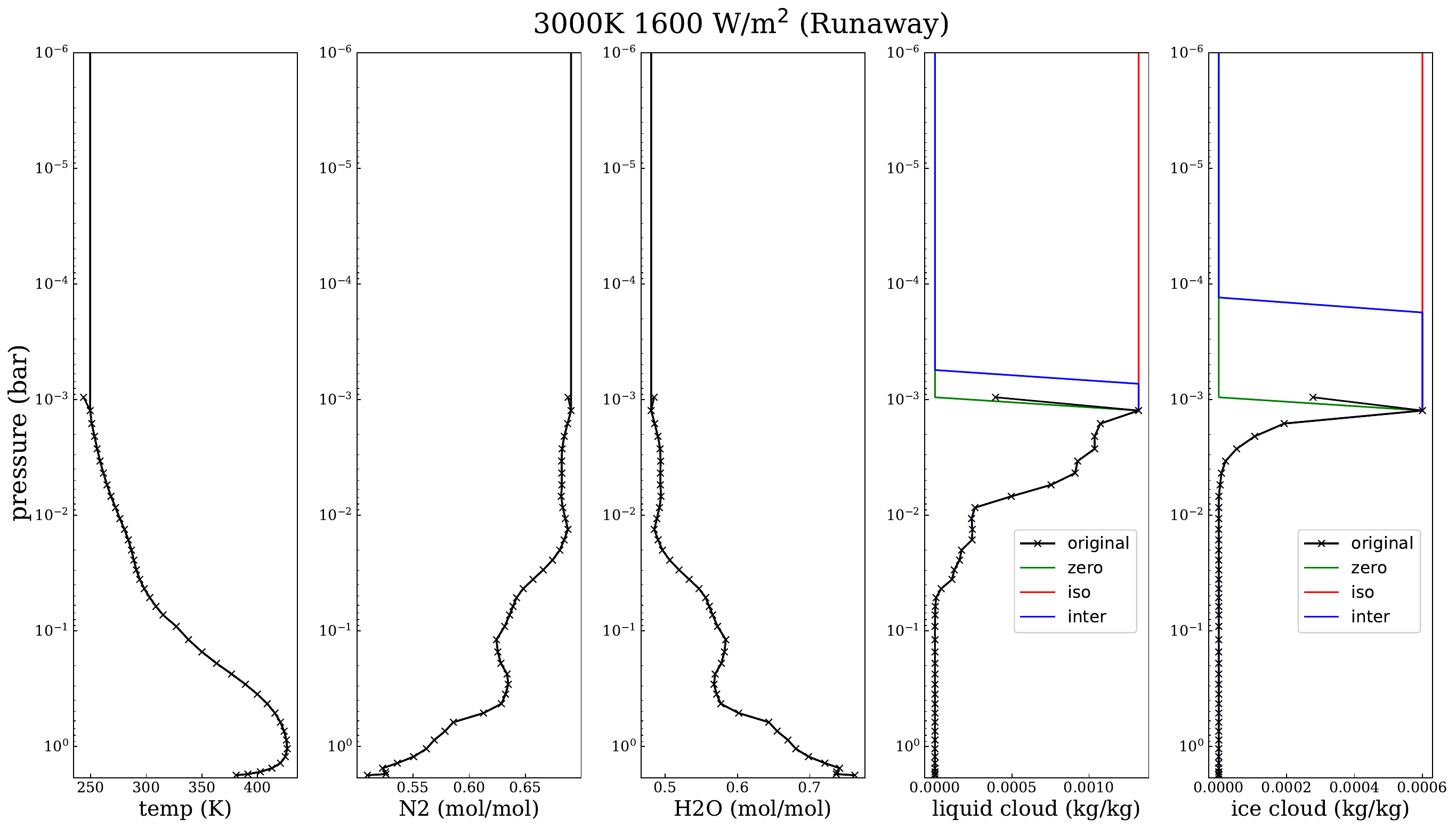}}\label{fig:runaway-profiles}
     \label{fig:profiles}
\end{figure}

As we discuss further in Section~\ref{sec:gcm}, it is necessary to extend the layers of the atmospheric profiles created by the GCM up to 1~$\micro$bar. The non-runaway planets are unaffected by additional layers because their cloud deck lies well below the original model top, as seen in the profiles displayed in Figure~\ref{fig:nonrunaway-profiles}. However, as Figure~\ref{fig:runaway-profiles} shows, for the incipient runaway cases, both the water and ice clouds accumulate at the model top (1~mbar), indicating that there are not enough layers in the GCM model profiles to properly capture the altitude and vertical extent of the cloud decks. Note that ice clouds are particularly problematic because they form at cold temperatures and thus occur higher in the atmosphere than liquid water clouds. We must therefore artificially add sufficient layers so that the model can reach 1~$\micro$bar, necessary for properly capturing the dynamic range of the transit spectra. To test how sensitive our spectra are to these low-pressure regions, we add the extra layers in three different ways:
\begin{itemize}
    \item Iso: Extend the upper layers by keeping all variables (temperature, H$_2$O and N$_2$ volume mixing ratios, liquid and ice cloud mass mixing ratios) constant, starting from the last viable point of the original model. Note that the top layer of GCMs can be unreliable due to artificial diffusion added to damp spurious wave reflection off the model top. 
    \item Zero: Extend the upper layers by keeping the temperature, H$_2$O, and N$_2$ volume mixing ratios as iso-values, but setting the liquid and ice cloud mass mixing ratios to zero. 
    \item Intermediate: Find the pressure at which the liquid and ice cloud mass mixing ratios reach half of its maximum abundance, determined from the original profile. This is treated as the bottom of the cloud deck. Ensure that the vertical extent of the cloud deck, starting from the bottom, reaches 1~dec in pressure. If the cloud deck in the original profile does not span 1~dec, we extend until 1~dec by setting the abundances as iso-values of the last viable point, and then set the remaining layers to zero once the 1~dec requirement is reached. 
\end{itemize}
The first two methods represent the limits of possible cloud abundances, while the intermediate profiles represent our best approximation based on our understanding of cloud profiles from our non-runaway cases as well as data on terrestrial planets in the Solar System. We include the unlikely extreme cases so we can constrain the impact of the highly uncertain low-pressure cloud regime from 1~mbar to 1~$\micro$bar on the resulting spectra. For the intermediate method, we have reason to assume a 1~dec vertical extent for the cloud decks, as some incipient runaway cases that do have a complete cloud deck below the model top span 1~dec of pressure. In addition, this is consistent with the cloud deck extent observed on Earth and Venus \citep{bezard:2007,taylor:2018}. We remind the reader that these approximations only make a difference for the planets in the incipient runaway regime. In Figures~\ref{fig:nonrunaway-profiles} and~\ref{fig:runaway-profiles}, one can see the profiles including the added layers using the three different methods listed above. For all of the above model top treatments, we keep the temperature constant because we make no temperature-dependent changes in the clouds or in the molecular abundance of water in the upper atmospheres. In general, transmission spectroscopy is insensitive to the shape of the pressure-temperature profile at these heights, so we deem it unimportant to determine a realistic mesospheric temperature profile. In addition, for all of the above methods, we first remove the lowest-pressure values of the original model to avoid any instabilities posed by the GCM while approaching the model top. 

\subsection{Determining Observable Targets for Future Observatories}
\label{sec:observatories}

Our ultimate goal for this study is to determine the observability of ocean Earths using transmission spectra with our current and future technological capabilities and limitations. As we discuss in Section~\ref{sec:spectral features}, the four major water vapor signatures in the infrared are the \SI{1.4}{\micro\metre}, the \SI{1.8}{\micro\metre}, the \SI{2.7}{\micro\metre}, and the \SI{6}{\micro\metre} features. All of these features are within the bandpasses of current and next-generation instrumentation dedicated to atmospheric characterization of exoplanets. The James Webb Space Telescope (JWST), estimated to launch in March 2021, will have spectral capability ranging from \SI{0.6}{\micro\metre} to \SI{28}{\micro\metre} with its NIRSpec (Near-Infrared Spectrograph), NIRCam (Near-Infrared Camera), and MIRI (Mid-Infrared Instrument) instruments, though it will most likely be limited to less than \SI{12}{\micro\metre} for transmission spectroscopy \citep{jwst:2014}. JWST would therefore be able to provide spectral coverage over all four of the primary water features in the planetary spectrum outlined in Figure~\ref{fig:3300}. 

In addition to JWST, there are several flagship space telescope mission concepts being examined for consideration by the upcoming Decadal Survey that will include spectroscopic capabilities for exoplanet characterization. If selected, the Large Ultraviolet Optical Infrared Surveyor (LUVOIR) would have capabilities for grism spectroscopy from \SI{0.2}{\micro\metre} to \SI{2.5}{\micro\metre} courtesy of the HDI (High-Definition Imager) instrument \citep{luvoir:2019}. As such, LUVOIR would be able to observe the 1.4 and \SI{1.8}{\micro\metre} features seen in Figure~\ref{fig:3300}. There are currently two mission architectures for LUVOIR, one with a 15-m mirror (LUVOIR Architecture A), and one with a $\sim$8-m mirror (LUVOIR Architecture B). We include both designs in our study. Another proposed mission concept, The Origins Space Telescope (OST), is currently designed to cover a wavelength range from \SI{2.8}{\micro\metre} to \SI{20}{\micro\metre} using a specially-designed instrument for transit spectroscopy called the the MISC (Mid-Infrared Imager, Spectrometer, Coronagraph) Transit Spectrometer \citep{origins}. OST would therefore have access to the large \SI{6}{\micro\metre} water vapor spectral feature. Each water vapor feature in the spectrum that we show in Figure~\ref{fig:3300} can thus be characterized by one or more upcoming/proposed telescopes (JWST, LUVOIR-A, LUVOIR-B, and OST). We note that it is not possible for both LUVOIR and OST to be selected. If either is chosen, the expected launch date would be in the mid-2030s.

\section{Results}
\label{sec:results}

\subsection{Spectral Features}
\label{sec:spectral features}
\begin{figure}
\centering
\resizebox{12cm}{!}{\includegraphics{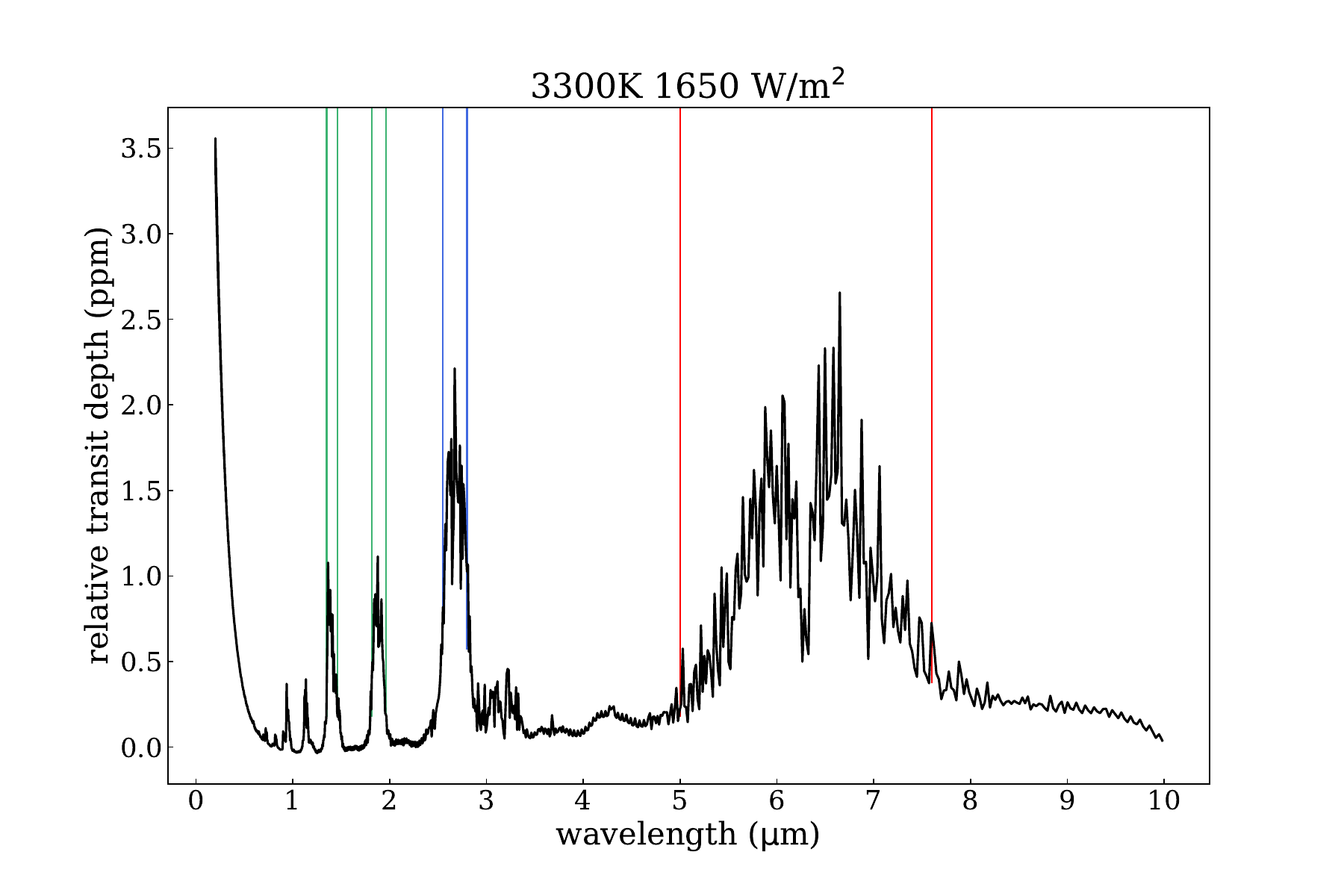}}
\caption{Example of a simulated spectrum for a planet in the non-runaway regime around a 3300~K star. The minimum value for the transit depth has been subtracted for ease of display, and the residual transit depth is given in parts per million of the overall stellar continuum. The vertical colored lines delineate the four different H$_2$O spectral features that we examined to determine exposure times for our four different telescope architectures (see Section~\ref{sec:observatories}).}
\label{fig:3300}
\end{figure}

In Figure~\ref{fig:3300}, we present a single case study of the PSG-synthesized spectrum for an ocean Earth receiving an incident flux of 1650 W/m$^2$ from a star with an effective temperature of 3300~K. As can be seen in Table~\ref{tab:sims}, this is a non-runaway case, and thus the spectrum in Figure~\ref{fig:3300} is independent of the treatment of the model top as described in Section~\ref{sec:modeltop}. We choose to discuss this GCM-modeled planet-star pair because \cite{ravi:2017} highlighted the spectral features for this specific case, as a representative example of a transmission spectrum for the same GCM results used here. As we will see in Section~\ref{sec:analysis}, the absorption features present in this spectrum are general for all pairs modeled by \cite{ravi:2017}, as the planetary atmosphere constituents are the same. In the synthesized spectrum, there are four major water vapor signatures in the infrared: the \SI{1.4}{\micro\metre}, the \SI{1.8}{\micro\metre}, the \SI{2.7}{\micro\metre}, and the \SI{6}{\micro\metre}, with the latter two being the most prominent. The small feature at \SI{4.15}{\micro\metre} visible for the 2600~K lower-flux cases is a N$_2$-N$_2$ collision-induced absorption feature, which was examined in detail by \cite{schwieterman:2015}. No features caused by other molecules are present, as the GCMs of \cite{ravi:2017} only include H$_2$O and N$_2$ in the atmospheres of the planets. The bulk of our analysis will focus on the water vapor features as they are indicators of potential habitability. 

\citet{ravi:2017} reported that the depth of H$_2$O spectral features would be observable due to the circulation of clouds around the substellar point for slow and synchronously rotating planets. For the specific scenario of an ocean Earth experiencing 1650 W/m$^2$ of flux from a 3300~K star, their simulated spectrum using the SMART radiative transfer model \citep{SMART:1996,SMART:1997} estimated a depth of $\sim$15~ppm for the \SI{6}{\micro\metre} feature. However, we correct this previous optimistic value by implementing three changes. We first recalculate the radius of the star. The radius that \citet{ravi:2017} used to simulate spectra was not consistent with the star's temperature. We adjust the radius from 0.137~R$_\odot$ to 0.301~R$_\odot$ using the Stefan–Boltzmann Law. This correction vastly reduces the average signal of the \SI{6}{\micro\metre} H$_2$O feature by $\sim$10~ppm. The second correction relates to the averaging scheme. Whereas \citet{ravi:2017} first averaged the profiles of the GCM results along the terminator to derive a single averaged profile and create a single simulated spectrum from this, we use the profile at each 4$\degree$ interval of latitude for a separate radiative transfer calculation, and then average all of the resulting spectra. This correction affects the spectral depth by adding $\sim$1~ppm to the average depth of the \SI{6}{\micro\metre} H$_2$O feature. Finally, we include the abundance of ice clouds from the model, which brings the average depth of the \SI{6}{\micro\metre} H$_2$O spectral feature down to $\sim$2~ppm (see Figure~\ref{fig:3300}). We note that these corrections affect the prospects for observability for these planets, but not the actual GCM results. We will continue to explore the spectra of the other planet-star pairs modeled by \cite{ravi:2017} before we predict how capable these telescopes will be in characterizing the atmospheres of these ocean-covered Earths. 

\subsection{Spectral Analysis for All Models}
\label{sec:analysis}

\begin{figure}
\centering
\begin{subfigure}
\centering
\includegraphics[totalheight=6.5cm]{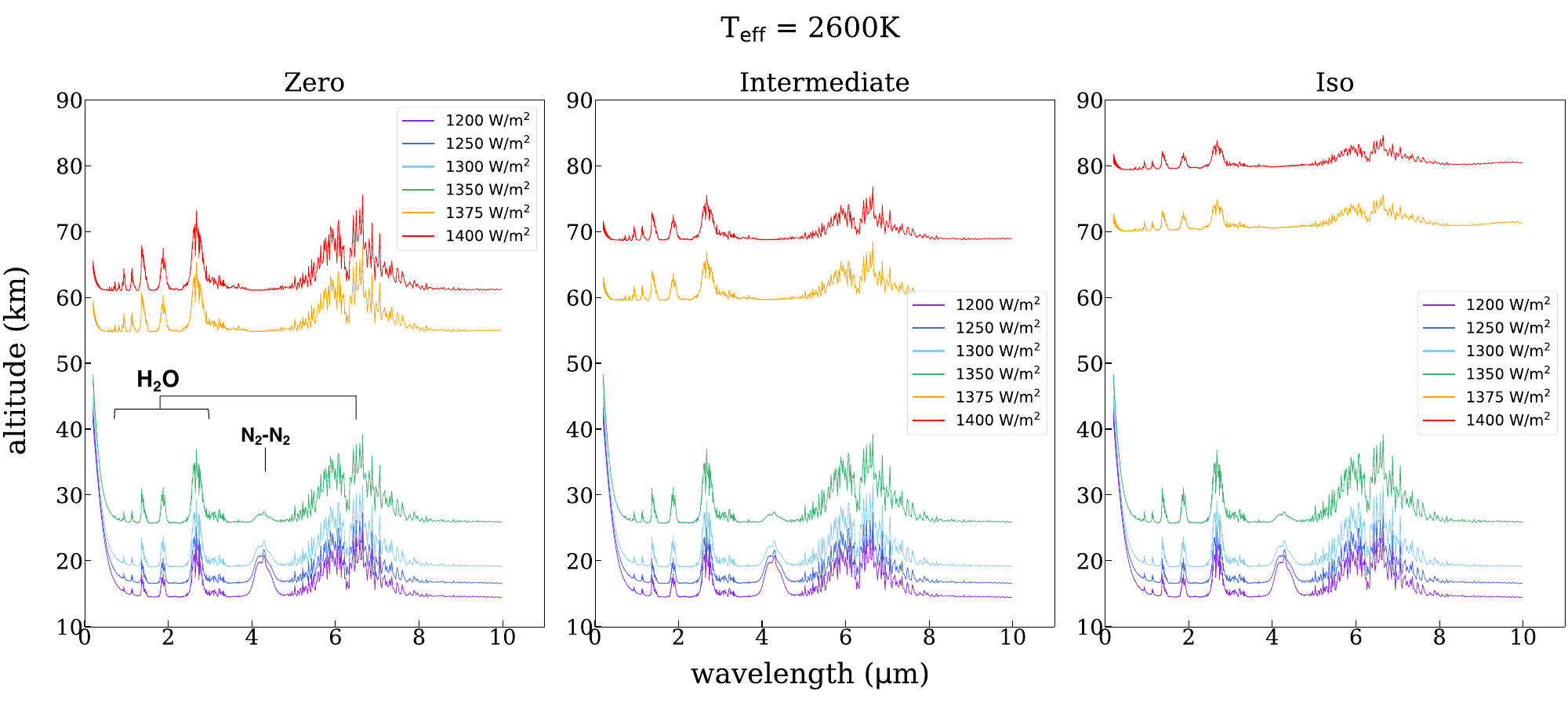}
\label{fig:2600rkm}
\end{subfigure}

\begin{subfigure}
\centering
\includegraphics[totalheight=6.5cm]{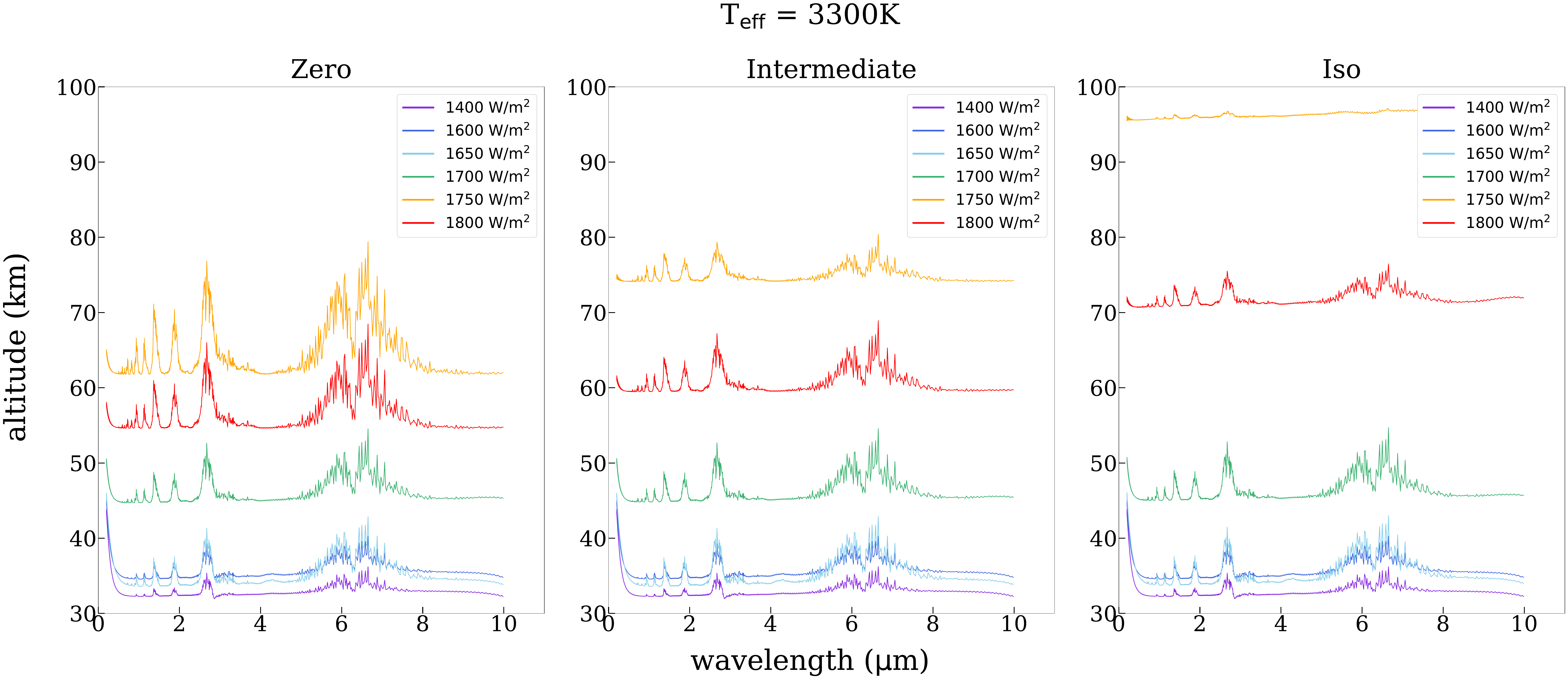}
\label{fig:3300rkm}
\end{subfigure}

\begin{subfigure}
\centering
\includegraphics[totalheight=6.5cm]{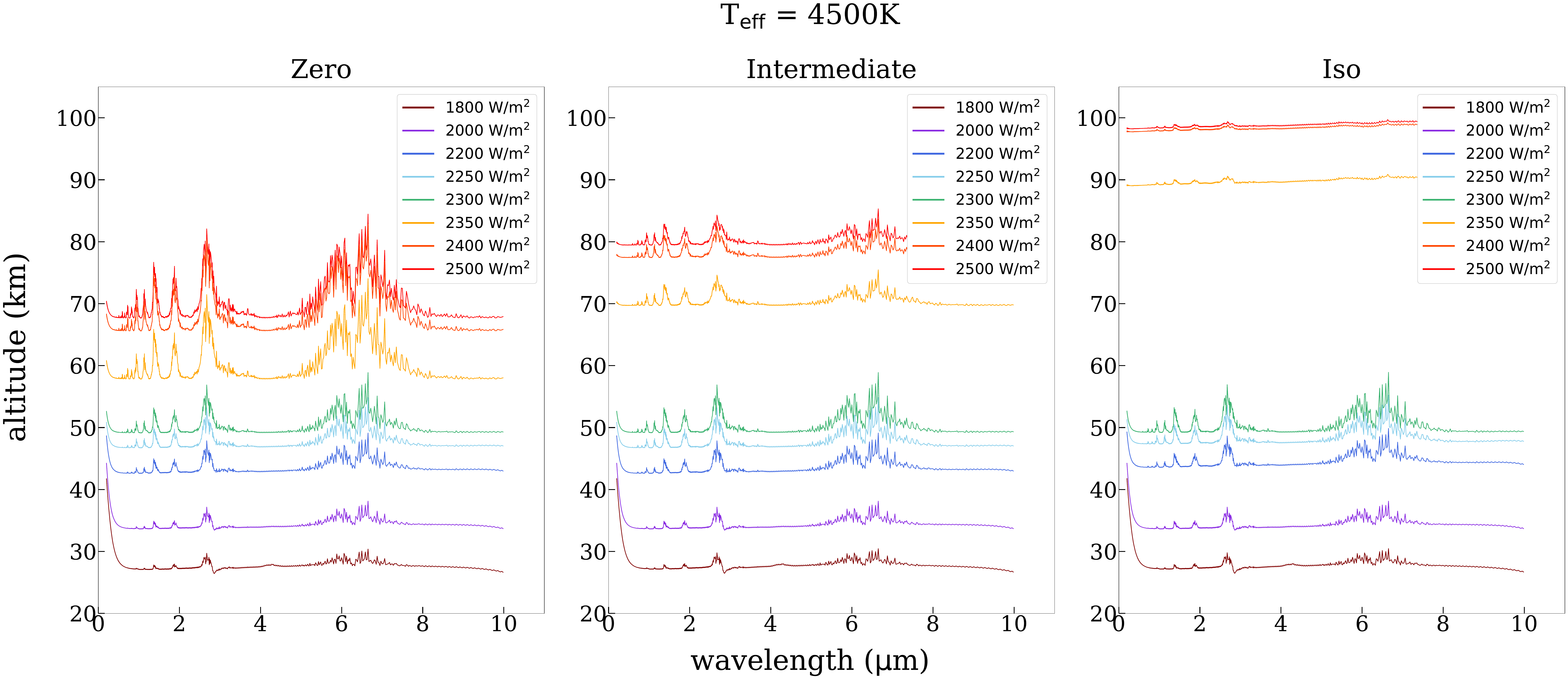}
\label{fig:4500rkm}
\end{subfigure}
\caption{Simulated spectra for every value of stellar irradiance for three different stellar effective temperatures. Results for three different model top assumptions (zero, intermediate and iso) are shown. The model top assumptions only significantly impact the incipient runaway regime, but result in significantly different results for the vertical extent of spectral features. Additionally, the altitude of the cloud deck rises significantly for incipient runaway cases.}
\label{fig:all-rkm}
\end{figure}

\begin{figure}
\centering
\includegraphics[width=\textwidth]{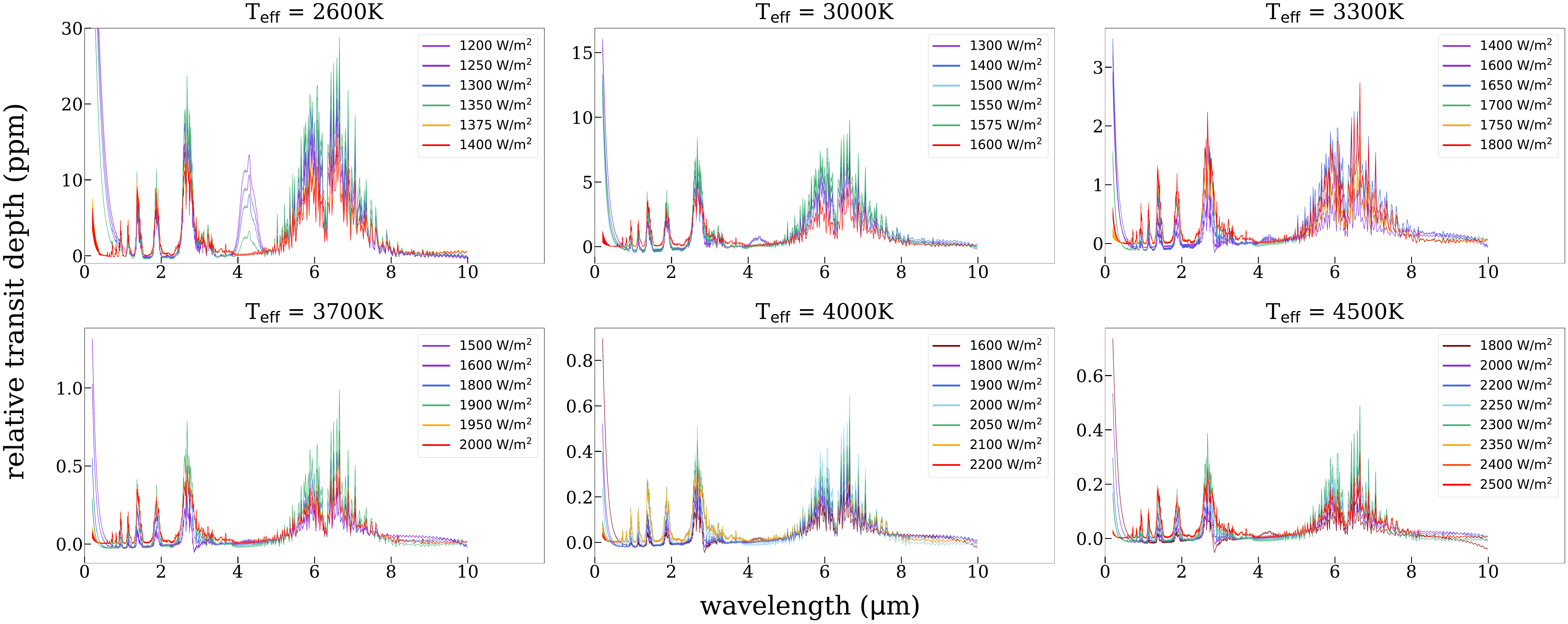}
\caption{The same simulations shown in Figure~\ref{fig:all-rkm}, but converted to relative transit depth in parts per million of the stellar continuum. The continuum transit depth has been subtracted. Both the Rayleigh scattering slope and the N$_2-$N$_2$ spectral feature at \SI{4.15}{\micro\metre} are only visible for the 2600~K simulations due to the low altitude of the cloud deck, but only H$_2$O spectral features are visible for later spectral types. The overall relative transit depths for all spectral features decline dramatically with increasing effective temperature, primarily due to the increasing stellar radius. Note that the y-axis scale is different for each different star.}
\label{fig:all-ppm}
\end{figure}

Using PSG, we synthesize the transmission spectra for all 39 planet-star pairs modeled by \citet{ravi:2017} (see Table~\ref{tab:sims}). The simulated spectroscopy results are displayed in Figures~\ref{fig:all-rkm} and \ref{fig:all-ppm}, in altitude and parts per million, respectively. Figure~\ref{fig:all-rkm} is divided into sub-graphs based on stellar temperature of the host star; we only show temperatures of 2600~K, 3300~K, and 4500~K for simplicity. Each sub-graph has three panels displaying the spectra using the iso-profiles, the zero-profiles, and the intermediate-profiles, with multiple spectra in each panel for the varying incident flux the planet receives. In Figure~\ref{fig:all-rkm}, the continuum (the relatively flat base of the spectrum where no spectral lines are present) represents how high the atmosphere's cloud deck lies above the planetary surface in kilometers. Therefore, the features visible in the presented spectra are solely due to the light from the star interacting with the molecules in the atmospheric layers above the cloud deck. Molecules below the cloud deck are inaccessible through transmission spectroscopy, because the cloud deck is optically thick. Within any panel representing stellar temperature in Figure~\ref{fig:all-rkm}, the cloud decks of the planets experiencing higher incident flux occur at higher altitudes than those of the planets with lower fluxes. This is consistent with expectations, since for the higher, the surface temperature of the planet the level at which the atmosphere has cooled enough to allow water to condense moves upward.

Figure~\ref{fig:all-rkm} also highlights that across all the sampled stellar temperatures, the spectra for the incipient runaway planets are highly dependent upon the nature of the upper atmospheric cloud decks. The features for the incipient runaway planets (typically the top one or two plotted lines in each panel, see Table~\ref{tab:sims}) shrink and the continuum rises as the assumed cloudiness increases from zero-profile to iso-profile. This is consistent with what we expect from the model top treatment seen in Figure~\ref{fig:runaway-profiles}. For the zero-profile case, many of the ice cloud profiles do not reach 1~dec of vertical extent, explaining the clarity with which we see the features (reminiscent of Figure~\ref{fig:3300}). For the intermediate- and iso-profiles, the spectral features for the incipient runaway planets are either diminutive or completely flat. In the temperature regimes explored here (planets up to $\sim$400~K surface temperatures), the upper atmosphere should remain cold enough to condense clouds, providing significant challenges for transit spectroscopy. Of course other atmospheric constituents, not included in our model presently, could modify the stratospheric temperature profile through either cooling (e.g. CO$_2$) or heating (e.g. O$_3$, absorbing aerosols) resulting in changes to the expected cloud profiles. While low-top GCMs are generally sufficient for understanding the climate generalities of warm moist planets, they are not yet sufficient for interpreting future transmission spectra from these worlds. An understanding of high altitude (low pressure) cloud and aerosol formation and persistence will be critical for interpreting observations of hot moist planets. 

Figures~\ref{fig:all-ppm} divides the spectra in panels for various stellar temperatures just based on the intermediate profiles. Figure~\ref{fig:all-ppm} shows in general that the planets orbiting lower-temperature stars have much stronger spectral features than those orbiting higher-temperature stars (up to $\sim$30~ppm compared to less than 1~ppm). The depths of spectral features for the planets orbiting 3700~K, 4000~K, and 4500~K stars are less than 1~ppm and too small to be observed by JWST. This relation is in fact the inverse of what was anticipated by \citet{ravi:2017}. Because the planets are synchronously rotating, the orbital period (equal to the rotation period), stellar mass, stellar luminosity and the incident flux on the planet are inextricably related through Kepler's third law (see Eq.(3) in \citet{ravi:2016}). As a result, the planets orbiting within the HZ of low-mass and low-luminosity stars (such as the 2600~K and 3000~K stars) in general rotate faster (or have shorter orbital periods) than those orbiting the hotter dwarf stars. \citet{ravi:2017} predicted that the slower-rotating planets (planets around 4000~K and 4500~K stars) would have relatively larger spectral features than their faster-rotating counterparts because slow rotating planets have elevated water vapor in their stratospheres at any given temperature due to the nature of their general circulation. However, we find in fact that the severe reduction in planet-to-star size ratio for warmer stars far outweighs the increase in stratospheric water vapor. 

\begin{figure}
\centering
\includegraphics[width=0.8\textwidth]{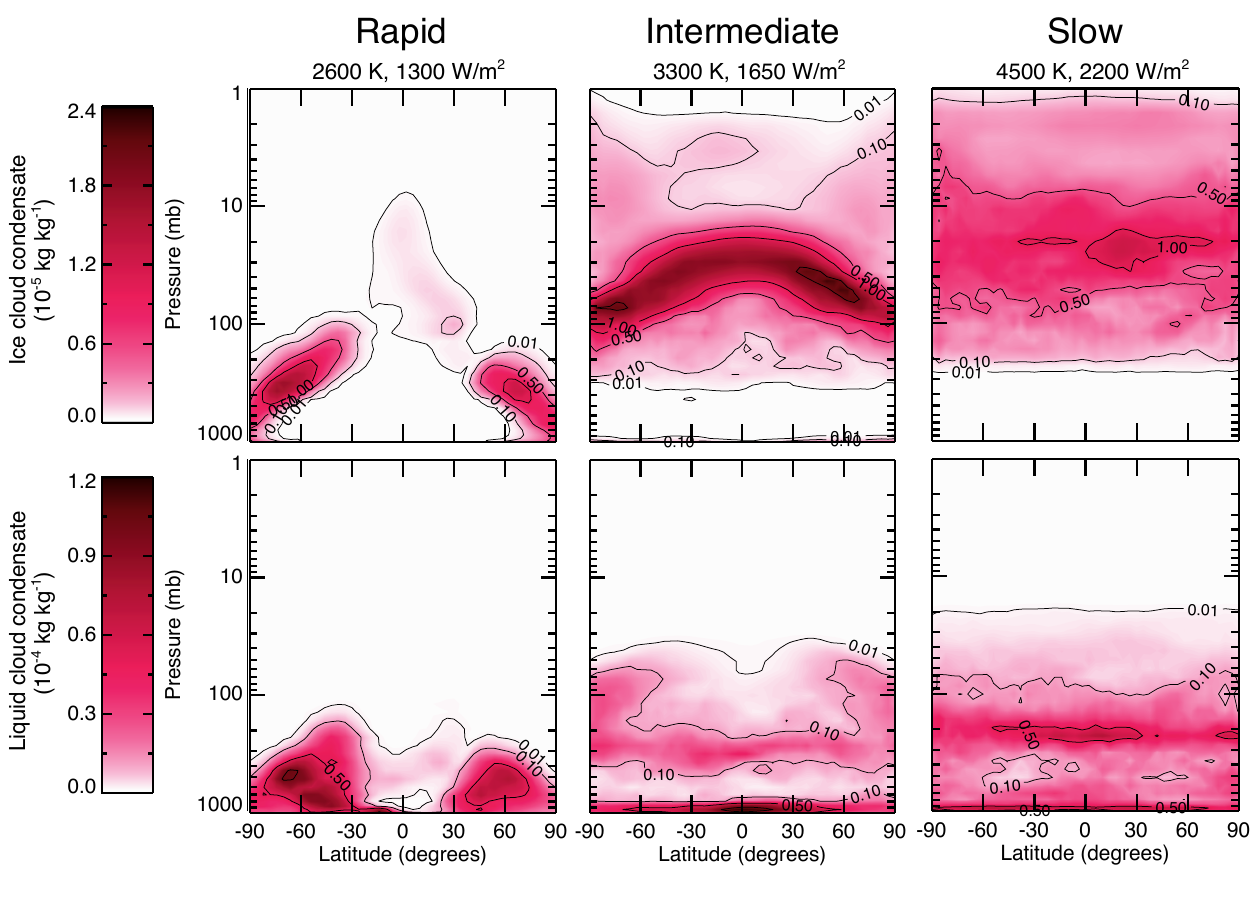}
\caption{Ice water cloud (top row) and liquid water cloud (bottom row) mixing ratios taken at the terminator, for temperate planets in rapid, intermediate, and slowly rotating regimes}. Cloud values shown are the average of east and west terminators. Despite each climate shown having similar mean surface temperatures ($\sim$280 K), changes to atmospheric circulation lower the cloud deck height as the planetary rotation rate increases.
\label{fig:circulation}
\end{figure}

\begin{figure}
\centering
\includegraphics[width=\textwidth]{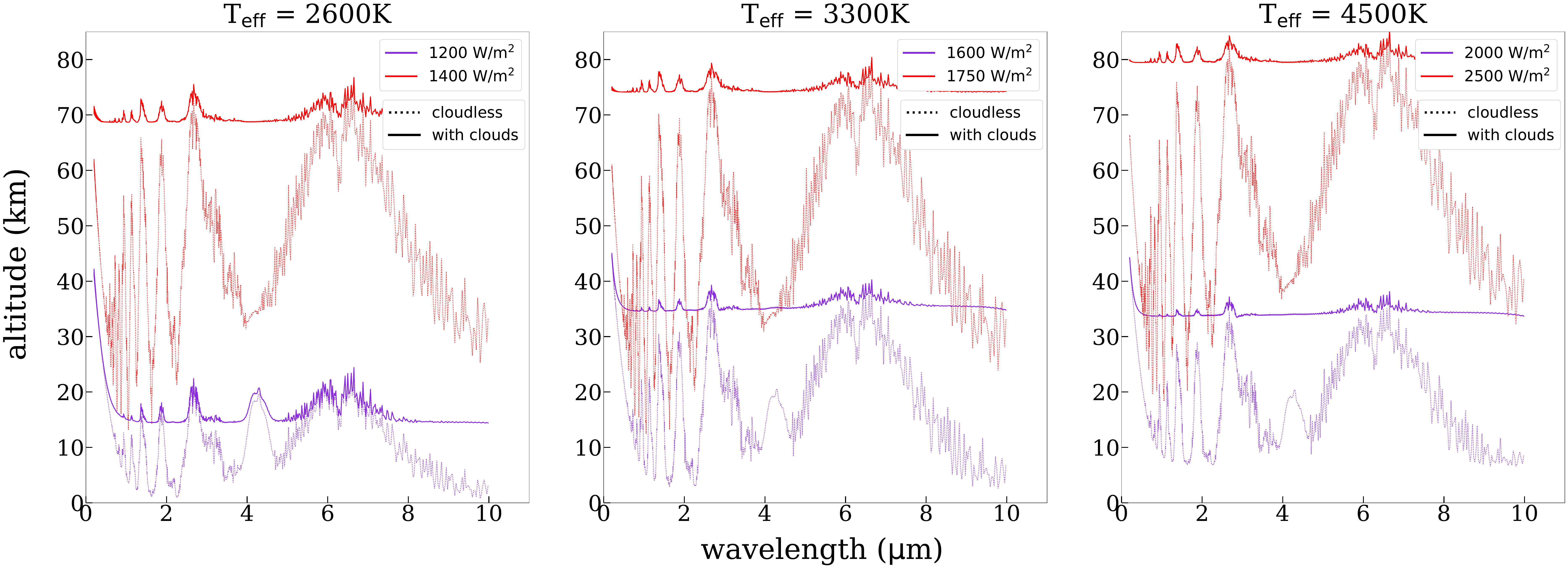}
\caption{Comparison of spectra for the same stellar temperatures as Figure~\ref{fig:circulation}, generated with cloud opacity (solid line) and without cloud opacity (dotted line). We show examples of both non-runaway (T$_{\mathrm{surf}} \approx 265$~K, purple) and incipient runaway (T$_{\mathrm{surf}} \approx 350$~K, red) models. The continuum opacity from clouds drastically diminishes the depth of absorption features compared with a cloud-free scenario.}
\label{fig:compareclouds}
\end{figure}

Changes to the atmospheric circulation also contribute to the deeper spectral features observed for the planets orbiting lower effective temperature stars. In Figure~\ref{fig:circulation}, we show the ice water cloud and liquid water cloud mixing ratios at the terminator for temperate planets around 2600~K, 3300~K, and 4500~K stars. The cloud abundances shown are the average of the east and west terminators. As discussed in \citet{haqq:2018} and elsewhere, as the planetary rotation rate increases, the atmospheric circulation transitions to different regimes. Circulation regimes have a substantial effect on the horizontal extent and location of clouds \citep{ravi:2017}, and also on the vertical extent of clouds. Relatively slowly rotating planets feature deep upwelling around the substellar point, resulting in thick, symmetric substellar clouds that reach high altitudes. However, as the planetary rotation rate increases, Coriolis forces support strengthened zonal flows, which cap upwelling motions in the atmosphere and shear the substellar clouds downstream. Thus, considering (approximately) identical mean surface temperatures, the cloud decks (both ice and liquid water) form progressively lower in the atmosphere as planetary rotation rate increases, as is seen in Figure~\ref{fig:circulation}. This is one contributing reason why the spectral features for the 2600~K stars in particular have a lower continuum height and are much deeper than those found for their higher stellar temperature counterparts. To highlight this effect, we display in Figure~\ref{fig:compareclouds} a comparison between spectra generated with clouds and without, for a selection of our models with corresponding surface temperatures.

In general, as Figure~\ref{fig:all-ppm} shows, even with significant stratospheric water vapor, transmission spectra of water-rich Earth-sized planets synchronously rotating around mid-K and early-M dwarf stars is exceedingly difficult. The planets orbiting the smallest star -- the star with an effective temperature of 2600~K -- have the most prominent spectral signals due to a combination of a high planet-to-star ratio and cloud asymmetry.

\subsection{Trends in Spectral Feature Depth}
\label{sec:trends}
In order to understand the relationship describing the different feature depths detailed in Section~\ref{sec:analysis} and Figure~\ref{fig:all-rkm}, we directly measure and compare the effective increase in planetary radius for each water vapor feature (\SI{1.4}{\micro\metre}, \SI{1.8}{\micro\metre}, \SI{2.7}{\micro\metre}, and \SI{6}{\micro\metre}) for all 39 available spectra. For each feature of each spectrum, we subtract the average value for the continuum from the average increase in planetary radius calculated within the spectral feature. We call this value $\Delta R$, or the change in the apparent radius of the planet (in kilometers) across a specific wavelength range. $\Delta R$ is simply the depth, or the strength, of a spectral feature. 

\begin{figure}
\centering
\includegraphics[width=\textwidth]{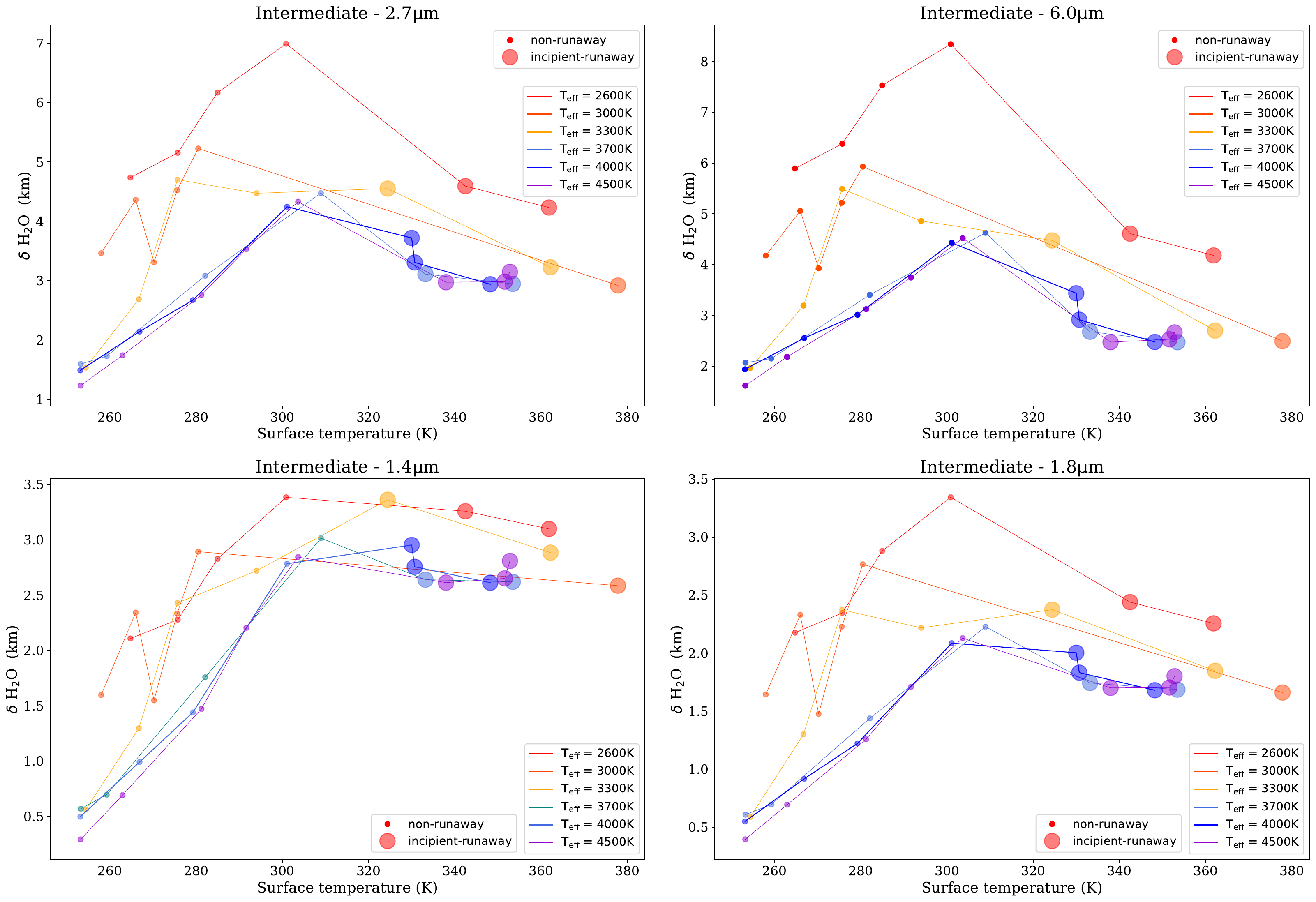}
\caption{For each H$_2$O spectral feature designated in Figure~\ref{fig:3300}, we plot the effective increase in planetary radius within the feature ($\Delta R$), which is independent of the stellar radius. $\Delta R$ generally rises with increasing surface temperature, as the scale height of the atmosphere increases, but then falls for incipient runaway planets due to the substantial increase in cloud deck altitude $-$ assuming an intermediate model top treatment. These trends are the same for stellar effective temperatures above 3700~K, but lower effective temperatures produce higher values, mostly due to a lower cloud deck altitude. }
\label{fig:delta-all}
\end{figure}

Figure~\ref{fig:delta-all} plots each simulated planet's $\Delta R$ value against its global surface temperature. Each panel in Figure~\ref{fig:delta-all} represents a different water vapor feature. For the incipient runaway cases, we plot results for the intermediate model top assumption; we mark these cases separate since they are strongly dependent on the model top assumptions chosen. Figure~\ref{fig:delta-all} shows that for the higher temperature host stars (3700~K, 4000~K, 4500~K), the feature depths generally increase with surface temperature until they reach an incipient runaway state, since the scale height $H=kT/mg$ is proportional to the temperature. However, when the runaway state begins, the lowest layers of the atmosphere become more and more opaque due to clouds, raising the continuum and decreasing the depth of each line. This is consistent with \citet{fujii:2017}, who report an increase in feature depth as the incident flux increases. 

Figure~\ref{fig:delta-all} shows that, in general, the transition between non-runaway and runaway regimes occurs when the planet experiences globally-averaged surface temperatures above $\sim$310~K. We can use the orbital parameters for simulated planets at this transition boundary to calculate an empirical relationship relating orbital semi-major axis to stellar temperature, in order to categorize discovered ocean Earths orbiting M stars into these two states:

\begin{equation}\label{eq:runaway}
a_{\textrm{transition}} = \num{-6.852e-11}\textrm{T}^{3} + \num{7.720e-07}\textrm{T}^{2} - \num{2.664e-03}\textrm{T} + 2.934
\end{equation}

in which the variable $a_{\textrm{transition}}$ is the semi major axis of the planet in AU and T is the stellar temperature in Kelvin. If $a < a_{\textrm{transition}}$, then the planet should be expected to be in a runaway greenhouse state. 

\begin{figure}
\centering
\includegraphics[width=0.6\textwidth]{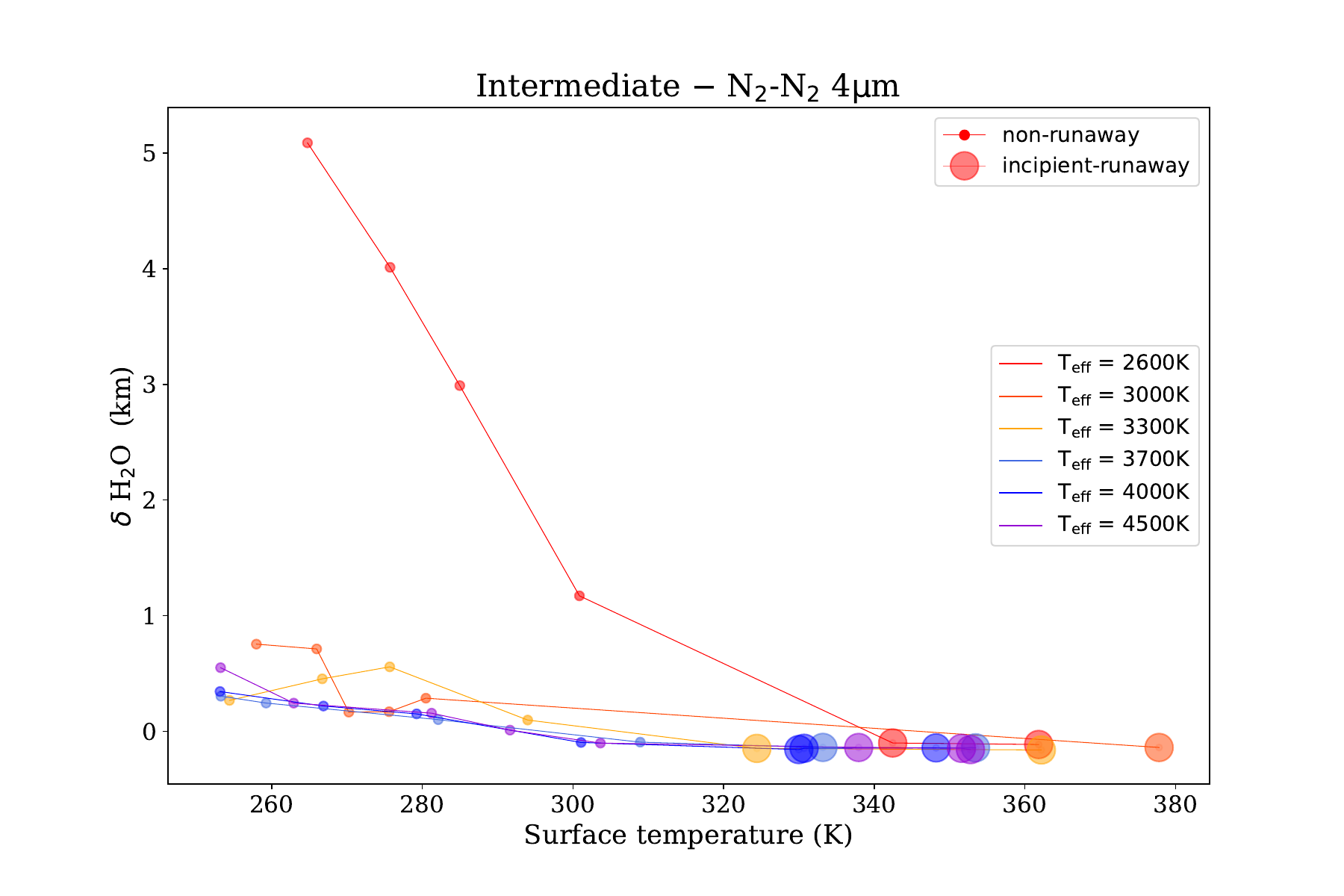}
\caption{The same as Figure~\ref{fig:delta-all}, but for the N$_2-$N$_2$ dimer spectral feature. The spectral absorption is only significant when the cloud deck is sufficiently low ($< 30$~km) to allow absorption by a significant amount of N$_2$, and is therefore only visible for cool atmospheres in the 2600~K case. }
\label{fig:delta-N2}
\end{figure}

Figure~\ref{fig:delta-N2} shows the $\Delta R$ value of the \SI{4}{\micro\metre} N$_2-$N$_2$ collision-induced absorption (CIA) feature against global surface temperature. It is clear that only the planets orbiting the 2600~K stars have noticeable dimer features. This is due to the dramatic difference of the cloud deck heights between the planets orbiting the 2600~K star and the planets orbiting higher temperature stars, as seen in Figure~\ref{fig:all-ppm}; the spectral feature achieves significant depth only when the cloud-based continuum is lower than 30~km, as see in Figure~\ref{fig:all-rkm}. The maximum depth of the spectral feature for the 3000~K models is $\sim$3~ppm, similar to results by \cite{schwieterman:2015} for a similar stellar type. However, for even cooler stars, the effective continuum occurs at even lower altitudes, and the lowest-flux models produce a signal up to 11~ppm. The impact of clouds on the intensity of CIAs is so strong because their opacity scales quadratically with pressure. They are only prominent at high pressures (low altitudes), and their intensity quickly drops as clouds mask those high pressures. At pressures above 1~bar, the feature saturates at the core, and only the weaker peripheral features contribute to the overall contrast, so the overall intensity of the dimer remains practically unchanged when the surface pressure is increased beyond 1~bar. As clouds mask the lower pressures, the intensity of the N$_2$ dimer drops substantially, becoming undetectable at pressures lower than 10~mbars ($>$30~km) as shown in Figure~\ref{fig:all-rkm}.

\subsection{Selecting TESS Targets}
\label{sec:tess}

We can utilize our calculations for the depth of H$_2$O spectral features to calculate how long it would take for future telescopes to detect water vapor on an ocean planet orbiting within the Habitable Zones of observable cool dwarf stars. To do this, we compile a target list of observable cool dwarfs from the TESS Input Catalog (TIC), since the TESS mission will largely dictate the available database of transiting exoplanet systems with terrestrial planets that the exoplanetary community will work with for the next few decades. 
 
TESS will observe 200,000 pre-selected stars at a two-minute cadence and take Full Frame Images (FFI) of all pixels every 30 minutes \citep{tess:2014}. Although TESS will observe an area 40 times larger than Kepler did, flux contamination will be more problematic for TESS as its pixels are 25 times larger \citep{tic:2019}. Due to this flux contamination, in combination with the inherent dimness of cool stars, it is unclear how many Earth-size planets TESS will discover around in our stellar temperature range of 2600~K $<\mathrm{T_{eff}} <$ 4500~K. However, even if TESS is not able to obtain sufficient sensitivity to detect small planets, augmentation through current and future ground-based and space-based observations may be able to achieve higher-quality light curves and detect additional planets.

One problem that the TIC may pose to our work is that dwarf stars may be misidentified as sub-giants. It is estimated that as many as 50\% of stars labelled as ``dwarfs" in the previous TIC version 7 were actually subgiants \citep{tic:2019}. However, thanks to the parallax measurements and photometry from the Gaia second data release, stellar radii in the TIC were revised in version 8, which greatly diminished the impact of this misidentification \citep{tic:2019}. This degeneracy will not affect our calculations, but merely re-classify some stellar targets in our target list as irrelevant (as seen in \citet{kane:2018}, for example). By using Gaia DR2 as the base, the number of stars in TICv8 that have estimated radii and effective temperatures increased by factors of 2 and 20, respectively. These catalog improvements work to our advantage when it comes to selecting relevant stars. 

From the TIC Version 8, the latest version at the time of publication, we select stars between the temperature range of 2600~K $< \mathrm{T_{eff}} <$ 4500~K. We then make additional cuts on our subset of stellar targets in order to maximize exposure time. We only consider stars with $\mathrm{R} < 0.5~\mathrm{R}_{\odot}$ to optimize the planet-star signal. In addition, we only select stars with a K$_{\textrm{mag}}$ brighter than 11. We choose this magnitude threshold so that we can maximize the photon flux of our stars to decrease our photon-noise limit, while still including ultra-cool dwarfs with small stellar radii similar to TRAPPIST-1 (K$_{\textrm{mag}} = 10.3$, \citet{gillon:2017}), the current benchmark for ultra-cool transiting systems. The combination of these three criteria result in a subset of 52,412 cool stars from the TESS Input Catalog. 

With some exceptions, most of the stars we are interested in for this study fall under the Cool Dwarf Catalog (CDC), a sub-catalog curated specially for the TIC \citep{muirhead:2018}. This list includes cool dwarf targets with $ V - J > 2.7$ and $\mathrm{T_{eff}} \lesssim 4000~\mathrm{K}$. \citet{muirhead:2018} determine effective temperatures for the cool dwarfs in their catalog by implementing the color-T$_{\mathrm{eff}}$ relations found in \citet{mann:2015,mann:2016}. However, \citet{muirhead:2018} do not apply this relation to stars with $\mathrm{T_{eff}} \lesssim 2700\mathrm{K}$ because the calibration sample used in \citet{mann:2015} does not include stars in this low-temperature regime. Therefore certain stars in the CDC do not have an assigned temperature, and thus did not survive our initial stellar effective temperature cut of 2600~K $< \mathrm{T_{eff}} <$ 4500~K. Because M dwarfs in this low-temperature regime (such as TRAPPIST-1) are of great interest to the community, we estimate the stellar temperatures for these stars ourselves using the discontinuous $\mathrm{T_{eff}}$-radius relation from \citet{rabus:2019}. \citet{rabus:2019} use high-precision interferometry measurements of M dwarfs to derive two separate empirical relations between $\mathrm{T_{eff}}$ and radius for stars with masses $\mathrm{M} < 0.23~\mathrm{M}_{\odot}$ and $\mathrm{M} \geq 0.23~\mathrm{M}_{\odot}$ respectively. For stars without $\mathrm{T_{eff}}$ but with $\mathrm{R} < 0.5~\mathrm{R}_{\odot}$ and K$_{\textrm{mag}} < 11$, we implement these relations to estimate effective temperatures. For stars with effective temperatures that lack radii but are marked as dwarfs in the TIC and have K$_{\textrm{mag}} < 11$, we implement the reverse of these relations to estimate radii. We only retain the stars with calculated radii of $\mathrm{R} < 0.5~\mathrm{R}_{\odot}$. These calculations add 9,101 new stars to our original dataset, bringing our total target list of M dwarfs up to 61,513 stars suitable for exposure time calculations (see Section~\ref{sec:calc-exptimes}). In our attached target list file, we flag which relation, if any, is used to calculate temperature or radius. 

We then precede to calculate exposure times for an ocean Earth orbiting around each star, as observed by either JWST, LUVOIR, or OST. One of the primary goals of the current era of transiting planet detection is to find targets for spectroscopy follow-up by the James Webb Space Telescope; however, we expand this scope to include all current concepts for flagship space telescopes with spectroscopy capabilities in the next decades (LUVOIR and the Origins Space Telescope, as detailed in Section~\ref{sec:observatories}). This list of exposure times for 61,513 stars will essentially be a ranking of which stars, if they had an ocean Earth, would be most amenable to a detection of water vapor by these observatories. 

\subsection{Calculating Exposure Times}
\label{sec:calc-exptimes}
To determine the detectability of an Earth-sized aquaplanet around one of the M stars in our target list, we compute how many hours it would take for each future observatory to detect a specific water vapor feature covered by the observatory's bandpass (Section~\ref{sec:observatories}). For this calculation, we must assume a $\Delta R$ using Figure~\ref{fig:delta-all}. We make an optimistic choice by choosing the maximum $\Delta R$ that a planet around a particular host star achieves before entering an incipient runaway state. For example, if a target star has an effective temperature of 2700~K, we assume that the planet reaches a $\Delta R$ of 8.3~km for the \SI{6}{\micro\metre} feature, as seen in Figure~\ref{fig:delta-all} for the 2600~K host star. Calculating exposure times for an incipient runaway planet will require a more robust treatment of the cloud properties for these planets, and is suggested as future work in Section~\ref{sec:discussion}.

We calculate for the exposure time $T_{exp}$ using the following equation:
\begin{equation}
   \label{eq:snr}
    \textrm{S/N} = \frac{\Delta\delta \cdot S_{t}\cdot\sqrt{T_{exp}}}{N_{t} \cdot \sqrt{2}}
\end{equation}
where $S_{t}$ is the photon count rate per unit time of the star, $N_{t}$ is the noise calculated per unit time, $\Delta\delta$ is the feature depth in ppm, and S/N is the signal-to-noise (set to 5 as our discretionary standard for detection).

$T_{exp}$ depends on $\Delta\delta$, which we can approximate using $\Delta R$ as introduced in Section~\ref{sec:trends}. To convert $\Delta R$, the planet's water vapor feature depth in kilometers, to $\Delta\delta$ (the same variable, in ppm), we use:
\begin{equation}
    \label{eq:ppm}
    \Delta\delta = \frac{(R_{p} + \Delta R)^2 - R_{p}}{R_{\star}^2}
\end{equation}

Our method for calculating the minimum exposure time necessary for a certain telescope to detect a designated water feature for a specific target star is as follows. We assign to each target star a model match based on its stellar effective temperature. For example, any target stars with effective temperatures between 2850~K and 3150~K are assigned to use the 3000~K 1575 W/m$^2$ simulation results (as stated above, we choose this flux because it is the flux at which the highest $\Delta R$ is reached before reaching an incipient runaway state). Once we have a model match for the target star, we take the corresponding model's spectrum in kilometers (as seen in Figure~\ref{fig:all-rkm}) and convert it from units of kilometers to ppm using the target star's radius from the TIC and Equation~\ref{eq:ppm}. 

We use PSG's instrument simulation module to simulate the signal and noise (both in photons) that a specific telescope will collect if observing a star of the brightness, size, and temperature of the target star. PSG uses the spectral radiance of the source, along with telescope and detector properties to compute the integrated flux and an associated noise contribution ($S_{t}$ and $N_{t}$). PSG's noise simulator incorporates background sky sources, photon noise from the source, and optical parameters of the telescope's instruments. 

We list the detector parameters we use with PSG for each telescope in Table~\ref{tab:noise}. For simplicity, we only use the instrument parameters for a single instrument on each observatory - otherwise, determining the optimal observing strategy for each of our 61,513 target stars would be prohibitively difficult, and would also make direct comparisons between individual targets more complex. In particular, we assume the parameters for the NIRSpec Prism instrument mode for JWST because it has access to most of the features (the \SI{1.4}{\micro\metre}, \SI{1.8}{\micro\metre}, and \SI{2.7}{\micro\metre} water vapor features). We do not calculate exposure times for the \SI{6}{\micro\metre} feature using the MIRI instrument because JWST is expected to be more sensitive in the near-infrared ($0.8 - $\SI{5.0}{\micro\metre}). We do not investigate in detail the effect of issues related to higher read-out noise for brighter targets, as this will only affect a small number of stars in our sub-catalog. In addition, we are aware that the stated brightness limit for the Prism using the Bright Object Time-Series Spectroscopy mode is $J\lesssim10$ \citep{beichman:2014}; however, the overall noise budget is dominated by photon noise or a noise floor, and therefore exposure time calculations assuming the NIRSpec Prism parameters will be roughly equivalent to what we would find for other higher-resolution modes for the JWST NIR instruments. The only exception would be the brightest targets in our target list, which may not be observable using any NIR instrument modes on JWST, but again we leave these in the target list for completeness. 

We calculate exposure time $T_{exp}$ using Equation~\ref{eq:snr}, assuming a required S/N of 5 for a detection. However, instead of simply subtracting the continuum from the maximum depth of the water feature, we optimize the bandwidth of the feature in order to obtain $\Delta\delta$. Starting with a small wavelength extent around the highest point, we compute the average depth of the feature within the extent and subtract it from the continuum. We then use this $\Delta\delta$ to calculate the exposure time with Equation~\ref{eq:snr}. We then increase the wavelength extent surrounding the feature incrementally and repeat, until we find the optimal $\Delta\delta$ value that yields the minimum exposure time possible for a specific target star-water feature-telescope pairing. 

\begin{table}
\centering
\caption{List of observatory and instrument parameters adopted for exposure time calculations. The values for JWST were taken from current documentation for the observatory, while the values for LUVOIR and OST were taken from the concept study reports.}
\label{tab:noise}
\setlength{\extrarowheight}{5pt}
\begin{tabular}{|c | c |c |c |c|}
\hline
 & JWST & LUVOIR-A & LUVOIR-B & OST \\
 \hline
 \hline
 Instrument & NIRSpec-Prism & HDI & HDI & MISC-TRA \\
 Wavelength & 0.7 - \SI{5.0}{\micro\metre} & 0.2 - \SI{2.2}{\micro\metre} & 0.2 - \SI{2.2}{\micro\metre} & 2.8 - \SI{20}{\micro\metre} \\ 
 Collecting diameter [m] & 5.64 & 14.05 & 7.46 & 5.64 \\ 
 Resolution [R] & 100 & 100 & 100 & 100 \\ 
 Read noise\footnote{\label{CCD}CCD}  [e$^-$] & 16.8 & 2.0 & 2.0 & -- \\ 
 Dark rate\footref{CCD} [e$^-$/s] & 0.005 & 0.001 & 0.001 & -- \\ 
 Sensitivity\footnote{NEP (Power Equivalent Noise Detector Model)} [W/$\sqrt{\text{Hz}}$] & -- & -- & -- & \num{2e-20} \\
 Total throughput & 0.4 & 0.1 - 0.342\footnote{\label{thru}Wavelength-dependent function} & 0.1 - 0.342\footref{thru} & 0.191 - 0.410\footref{thru} \\ 
 Emissivity & 0.1 & 0.1 & 0.1 & 0.1 \\ 
 Temperature [K] & 50 & 270 & 270 & 4.5 \\ 
\hline

\hline
\end{tabular}
\end{table}

The above calculation methodology assumes that there is no absolute noise floor imposed on the observations. The noise floor introduced by a telescope is a crucial factor to consider \citep{greene:2016,fauchez:2019}, especially with the low spectral depths seen in Figure~\ref{fig:all-ppm}. We re-calculate exposure times with a 1, 3, and 5~ppm noise floor. Note that for each noise floor calculation, we still assume a S/N of 5, or a 5-$\sigma$ detection. To incorporate the noise floor at every wavelength extent in the optimization scheme, we also calculate $T_{floor}$, or the maximum exposure time a telescope can observe for before the photon noise reaches the noise floor; at this point, any signal from the star is drowned out. $T_{floor}$ is computed using the following equation: 
\begin{equation}
    \label{eq:noise-floor}
    T_{floor} = \frac{2\cdot N_{t}^2}{N_{ppm}^2\cdot S_{t}^2} 
\end{equation}
where $N_{ppm}$ is the noise floor ($1e^{-6}$, $3e^{-6}$, or $5e^{-6}$). If $T_{floor}$ is less than $T_{exp}$, then detection of the feature in that bandwidth is not possible due to the noise floor. At each wavelength extent, we calculate both $T_{exp}$ and $T_{floor}$, and find the maximum wavelength range at which $T_{exp}$ is less than $T_{floor}$ (if at all). 

\subsection{Exposure Time Results}
\label{sec:results-exptimes} 
\begin{figure}
\centering
\includegraphics[width=\linewidth]{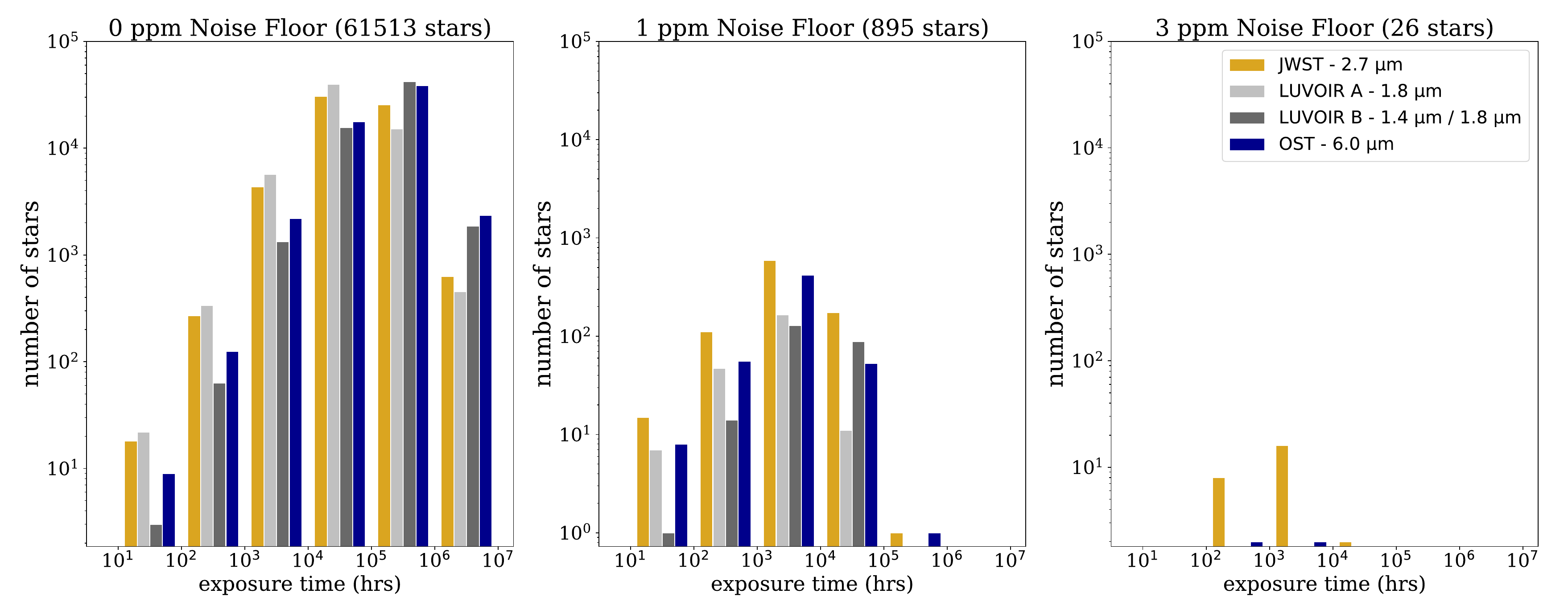}
\caption{Histograms of the number of stars which would allow for a 5-$\sigma$ detection of water, assuming different values for the absolute noise floor achievable. Exposure times are binned into decs in log space, with different colors representing different observatories. For LUVOIR-A and -B, the feature that yields the shortest exposure time is chosen for that star. With no noise floor, 20 stars could provide a detectable signal with JWST in less than 100 hours; however, for a noise floor of 3ppm, only 26 stars total could provide a detectable signal.}
\label{fig:exp}
\end{figure}

We present the results from our exposure time calculations in Figure~\ref{fig:exp}, which compares the computed exposure times of JWST, LUVOIR-A, LUVOIR-B, and OST using the assigned feature of each telescope that yields the lowest exposure times. As stated in Section~\ref{sec:calc-exptimes}, we do not exclude any targets based on brightness and saturation limits, since we want to compare the same target list across observatories and avoid delving too deeply into the subtleties of observing limitations. Although JWST can characterize three of the water vapor features using the NIRSpec Prism mode (\SI{2.7}{\micro\metre}, \SI{1.4}{\micro\metre}, \SI{1.8}{\micro\metre}), it is the \SI{2.7}{\micro\metre} feature that provides a detection in the shortest exposure time. On the other hand, LUVOIR-A and -B can see both the \SI{1.4}{\micro\metre} and \SI{1.8}{\micro\metre} features, and the best feature for which these telescopes require shorter exposure times to achieve a S/N of 5 depends on the temperature of the star. For this reason, for LUVOIR-A and LUVOIR-B in the first panel of Figure~\ref{fig:exp}, we only consider the shortest exposure time for each star regardless of which feature yields it. As detailed in Section~\ref{sec:observatories}, OST can only observe the \SI{6}{\micro\metre} H$_2$O feature. Each histogram in Figure~\ref{fig:exp} is binned by factors of 10 in exposure time; for example, the first bin includes stars that would require an exposure time of between 100 and 1000 hours. It is clear from the first panel in Figure~\ref{fig:exp} that the majority of the stars in our target list have infeasible exposure times, with the average exposure time for JWST being on the order of 10$^5$ hours. 

Each panel in Figure~\ref{fig:exp} contains a different noise floor consideration. Assuming no noise floor (0~ppm) with the method described in Section~\ref{sec:calc-exptimes}, LUVOIR-A and JWST could characterize the most stars in the shorter-exposure time regime; we list the number of stars that each telescope could observe under 1000 hours and under 300 hours assuming no noise floor in Table~\ref{tab:noisefloor}. When accounting for a 1~ppm noise floor, the number of stars whose planetary spectra do not get drowned out by the noise floor decreases drastically. For LUVOIR-A and -B, only exposure times using the \SI{1.8}{\micro\metre} feature survive the noise floor. When implementing a 3~ppm noise floor, the number of stars both LUVOIR-A and -B can characterize reach zero, since the depth of the \SI{1.8}{\micro\metre} feature never rises above 15~ppm for any stars. Only JWST and OST could detect the spectra of a handful of planets. Finally, we test a 5~ppm noise floor, in which no telescope can characterize the spectrum of a planet around any of our selected TESS stars. The numbers of stars that survive different noise floor considerations for each telescope are shown in Table~\ref{tab:noisefloor}. 

Returning to Equation~\ref{eq:snr}, we note that since $\Delta R$ and $R_p$ are constant for each calculation, and $N_{t} \approx \sqrt{S_{t}}$:
\begin{equation}
    \label{eq:prop}
    T \propto \frac{R_\star ^4}{S_{t}}
\end{equation} 

In other words, stars with smaller radii and brighter magnitudes (i.e, more photons) yield the shortest exposure times. It is these types of stars that will prove to be the best targets for characterizing ocean Earths. The explicit ranking of our stellar targets based on shortest exposure times is included as a supplementary digital file to the paper. The top 20 stars with the best S/N per hour of exposure time are shown in Table~\ref{tab:beststars} along with their stellar parameters; many of these stars would require a very low noise floor or would saturate all of the JWST NIR instruments, but we include them as our list of the most ideal targets if instrument/observatory capabilities allow it. The 26 stars that remain detectable with a 3~ppm noise floor are listed in Table~\ref{tab:3ppmstars}. These are all relatively dim stars, and therefore should all be observable by JWST as well as future observatories. 

\begin{figure}
\centering
\includegraphics[width=0.8\textwidth]{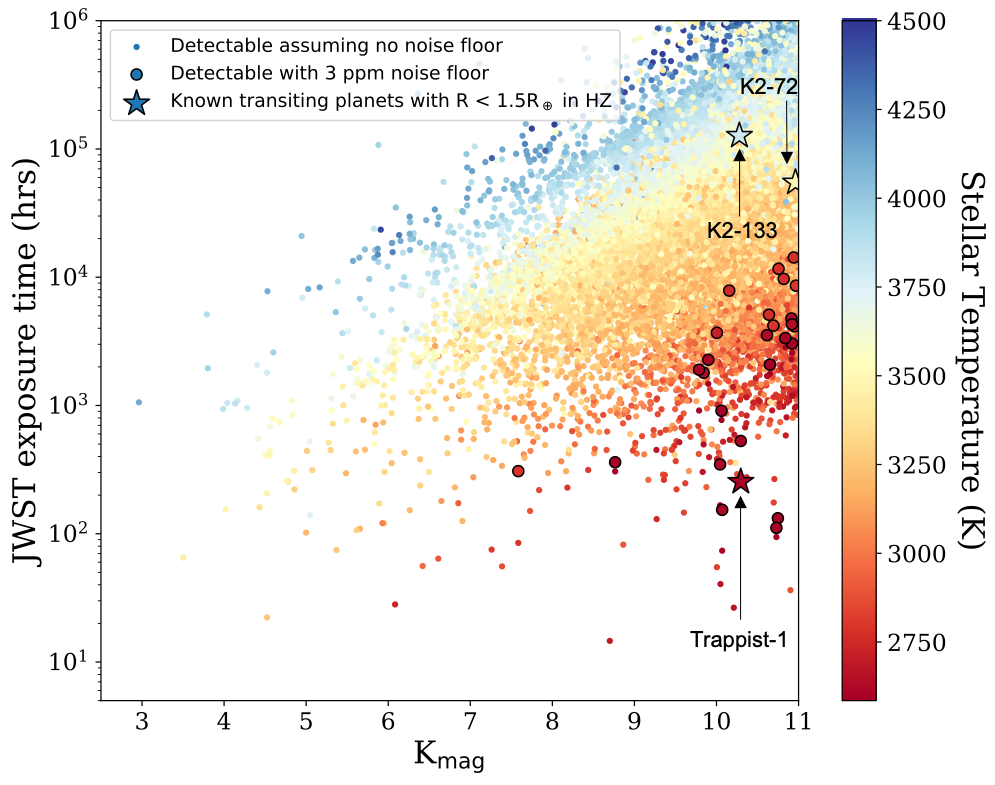}
\caption{Exposure time vs stellar K-band magnitude for all the TIC targets with K$_{\textrm{mag}}<$ 11 and total exposure time $<$ 10$^6$ hours, assuming photon-noise limited performance (no noise floor). Generally exposure time decreases with both stellar temperature and K$_{\textrm{mag}}$, as expected; targets with the lowest exposure times are listed in Table~\ref{tab:beststars}. If we assume an absolute noise floor of 3~ppm, only a small subset of the smallest stars remain detectable. Additionally, the three known Earth-sized planets in the Habitable Zones of their parent stars that have exposure times below 10$^6$ hours are indicated by star shapes. Only TRAPPIST-1e would be detectable with a 3~ppm noise floor.}
\label{fig:kmagplot}
\end{figure}

To unpack the trends in exposure time, we present Figure~\ref{fig:kmagplot}, where we plot each target star's exposure time for JWST $-$ with no noise floor $-$ against its magnitude and temperature. This figure shows that in general the stars with the lowest, most realistic exposure times are the brighter and cooler ones. This agrees with the relation provided by Equation~\ref{eq:prop}. The lower the stellar temperature, the smaller the stellar radius, leading to a shorter exposure time. The higher the K$_{\textrm{mag}}$, the dimmer the star, the longer the exposure time. The starred points in Figure~\ref{fig:kmagplot} represent stars that we already know to have a terrestrial planet in the HZ. We only label stars with planets with $\mathrm{R} < 1.5~\mathrm{R}_\odot$ that are at incident fluxes comparable to those in \citet{ravi:2017}, listed in Table~\ref{tab:sims}. Note that we do not mark LHS 1140b, due to its recent radius correction using Gaia DR2 data \citep{kane:2018}. Figure~\ref{fig:kmagplot} also marks the 26 stars that survive the 3~ppm noise floor. Note that these stars do not necessarily correlate to lower exposure times. These stars are the coolest, with effective temperatures $\sim$2600~K. As shown in Figure~\ref{fig:all-ppm}, only planets orbiting stars with T$_{\textrm{eff}} < $ 2600~K would reach spectral features with a depth of $\ge 15-20$~ppm. However, as shown in Table~\ref{tab:noise}, no Earth-sized aquaplanet around any of our target stars would be detectable using any telescope if we can only achieve a noise floor of 5~ppm. This is because not even the planets orbiting the 2600~K stars in Figure~\ref{fig:all-ppm} reach 25~ppm, the minimum feature depth required to still be visible with a 5-$\sigma$ detection and a 5-ppm noise floor. We analyze the implications of the noise floor and each telescope's prospects for realistic characterization in Section~\ref{sec:discussion}.

\begin{table}
\centering
\caption{Number of stars that would be detectable, either assuming a maximum exposure time with no noise floor or assuming that the stochastic noise floor is limited to a 1-sigma noise envelope of 1~ppm, 3~ppm, or 5~ppm.}
\label{tab:noisefloor}
\setlength{\extrarowheight}{5pt}
\begin{tabular}{|c | c|c|c|c|c|}
\hline
 & \multicolumn{5}{c|}{No. of stars detectable} \\
\hline
 & 0 ppm & 0 ppm & 1 ppm & 3 ppm & 5 ppm \\ Telescope & $T_{exp}<$ 1000 hrs & $T_{exp}<$ 300 hrs & Tot. & Tot. & Tot. \\
\hline
JWST, \SI{2.7}{\micro\metre} & 329 & 60 & 895 & 26 & 0 \\
OST, \SI{6.0}{\micro\metre} & 289 & 33 & 535 & 4 & 0 \\ 
LUVOIR-A, \SI{1.8}{\micro\metre} & 134 & 63 & 231 & 0 & 0 \\
LUVOIR-B, \SI{1.8}{\micro\metre} & 60 & 15 & 231 & 0 & 0 \\
\hline
\end{tabular}
\end{table}

\begin{table}
\centering
\caption{Ranked list of the targets from the TESS Input Catalog with the best prospects for detecting water vapor, assuming photon-limited performance and no brightness or other limits for the observatory. The stellar parameters are taken from the TIC when available, and calculated using relations from \citet{rabus:2019} when values are missing from the TIC. It is clear that the smallest stars are the best targets, along with several of the nearest and brightest mid-M stars.}
\label{tab:beststars}
\setlength{\extrarowheight}{5pt}
\begin{tabular}{|c|c|c|c|c|c|c|}
\hline
 & Identifier & flag\footnote{TRML $\equiv$ TIC lists stellar temperature, radius, mass, and luminosity} & K mag & Radius [R$_\odot$] & Temp [K] \\
\hline
1 & G 58-18A & RML & 8.7 & 0.129 & 2699\\
2 & Proxima Centauri & RM & 4.384 & 0.153719 & 3000\footnote{The radius-temperature relation from \citet{rabus:2019} underestimates temperature for Proxima Centauri. We override the temperature with 3000~K, an approximation from \citet{proxcen:2012}.}\\
3 & Barnard's Star & TRML & 4.524 & 0.193525 & 3259\\
4 & 2MASS J08262921+0424089 & RML & 10.211 & 0.124 & 2666\\
5 & Wolf 359 & RM & 6.084 & 0.135286 & 2741\\
6 & G 99-10B & TRML & 10.901 & 0.132256 & 2880 \\
7 & 2MASS J03293895+4441055 & RML & 10.048 & 0.124 & 2666\\
8 & LSPM J1016+3925 & RML & 10.005 & 0.149 & 2832\\ 
9 & G 208-45 & TRML & 7.387 & 0.142116 & 2865 \\ 
10 & YZ Ceti & RM & 6.42 & 0.167787 & 2957\\ 
11 & GJ 1061 & TRML & 6.61 & 0.155739 & 2905\\ 
12 & LP 938-71\footnote{Brown dwarf} & RM & 10.069 & 0.115627 & 2610\\
13 & Ross 154 & TRML & 5.37 & 0.210477 & 3261\\ 
14 & G 51-15 & TRML & 7.26 & 0.124013 & 2814\\ 
15 & 2MASS J04173685-2419503 & RML & 8.864 & 0.159 & 2898\\ 
16 & Teegarden's Star & TRML & 7.585 & 0.118649 & 2790 \\ 
17 & 2MASS J15402966-2613422 & RM & 10.73 & 0.113881 & 2598\\
18 & HD 173740 & RM & 5.0 & 0.278799 & 3213\\ 
19 & HD 153026B & RML & 5.614 & 0.218 & 3291\\ 
20 & Ross 128 & TRML & 5.654 & 0.209628 & 3163\\

\hline
\end{tabular}
\end{table}

\begin{table}
\centering
\caption{List of the targets from the TESS Input Catalog that are detectable with a 3~ppm noise floor using JWST, ranked by lowest exposure time. The stellar parameters are taken from the TIC when available, and calculated using relations from \citet{rabus:2019} when values are missing from the TIC. Note that many of these stars do not have temperatures in the TIC. Only the smallest, coolest stars remain detectable, but all should be observable with one or more of the JWST NIR instruments.}
\label{tab:3ppmstars}
\setlength{\extrarowheight}{5pt}
\begin{tabular}{|c|c|c|c|c|c|c|}
\hline
 & Identifier & JWST [hrs] & flag\footnote{TRML $\equiv$ TIC lists stellar temperature, radius, mass, and luminosity} & K mag & Radius [R$_\odot$] & Temp [K] \\
\hline
1 & 2MASS J15402966-2613422 & 111.181 & RM & 10.73 & 0.113881 & 2598 \\
2 & 2MASS J04133542+3127111 & 132.096 & RML & 10.748 & 0.113 & 2593 \\
3 & LP 938-71\footnote{\label{browndwarf}Brown dwarf} & 153.572 & RM & 10.069 & 0.115627 & 2610 \\
4 & Teegarden's Star & 307.744 & TRML & 7.585 & 0.118649 & 2790 \\
5 & LSR J0539+4038 & 347.351 & RM & 10.044 & 0.11641 & 2615 \\
6 & VB 10 & 360.890 & RM & 8.765 & 0.113185 & 2594 \\
7 & TRAPPIST-1 & 528.936 & RM & 10.296 & 0.114827 & 2605 \\
8 & LSPM J2044+1517 & 909.158 & RM & 10.061 & 0.111911 & 2585 \\ 
9 & LP 760-3 & 1801.495 & RML & 9.843 & 0.121 & 2646 \\ 
10 & LP 98-79 & 1901.540 & RM & 9.788 & 0.115487 & 2609 \\ 
11 & 2MASS J23312174-2749500 & 2083.498 & RM & 10.651 & 0.111914 & 2585 \\ 
12 & 2MASS J04195212+4233304 & 2270.569 & RM & 9.9 & 0.11416 & 2600 \\
13 & LP 135-272 & 3038.508 & RM & 10.922 & 0.113564 & 2596 \\ 
14 & 2MASS J07140394+3702459\footref{browndwarf} & 3351.862 & RM & 10.838 & 0.11384 & 2598 \\ 
15 & LP 326-21 & 3534.983 & RM & 10.616 & 0.116388 & 2615 \\ 
16 & LP 335-12 & 3688.572 & TRML & 10.005 & 0.121203 & 2722 \\ 
17 & 2MASS J14230252+5146303 & 4145.738 & TRML & 10.95 & 0.116383 & 2758 \\
18 & LP 535-12 & 4196.176 & TRML & 10.693 & 0.11867 & 2792 \\ 
19 & LP 441-34 & 4297.327 & RM & 10.919 & 0.119315 & 2635 \\ 
20 & 2MASS J21272531+5553150 & 4767.612 & RM & 10.913 & 0.117323 & 2621 \\
21 & LP 412-31 & 5102.946 & TRML & 10.639 & 0.115117 & 2766 \\
22 & 2MASS J15242475+2925318\footref{browndwarf} & 7866.570 & TRML & 10.155 & 0.121363 & 2755 \\
23 & 2MASS J17415439+0940537 & 8594.149 & TRML & 10.968 & 0.11895 & 2802 \\
24 & 2MASS J00251602+5422547 & 9710.488 & TRML & 10.819 & 0.11805 & 2772 \\
25 & LP 485-17\footref{browndwarf} & 11638.093 & TRML & 10.756 & 0.119322 & 2772 \\
26 & LP 423-14 & 14252.130 & TRML & 10.942 & 0.121019 & 2768 \\

\hline
\end{tabular}
\end{table}

\section{Discussion}
\label{sec:discussion}
\subsection{Comparison with Previous Work}
\label{sec:previouswork}
A large number of studies investigate Earth-like planets in the HZ; here, we constrain our discussion to previous work done on aquaplanets in particular or using 3-D GCMs for modeling water-rich worlds. Our most important result, that clouds produce the dominant impact on transit spectral characterization for slow and synchronous rotating planets, compares well with other GCM studies. For example, \citet{komacek:2020} have used ExoCAM to simulate planets synchronously rotating around M dwarfs with effective temperatures of 2600~K. They study the impact that clouds have on the detection of water vapor spectral features for planets with various rotation periods, surface pressures, planetary radii, and surface radii. The maximum depth of water vapor features for any of their simulated planets around a star with $\mathrm{T_{eff}} = $ 2600~K is 20~ppm, consistent with our results for planets orbiting stars of the same temperature. \citet{komacek:2020} also predict that for an Earth-sized planet with a rotation period of 4.11~days around a 2600~K star (similar to our planet with an incident flux of 1350~W/m$^2$ around a 2600~K star), it would take JWST NIRSpec/Prism $\sim$320 transits to detect its water features, without a noise floor. This is comparable to our calculation of $\sim$254 hours for a JWST characterization of a similar planet around TRAPPIST-1, without a noise floor (Section~\ref{sec:implications}). In addition, \citet{pidhorodetska:2020} have simulated TRAPPIST-1e observations by LUVOIR, HabEx, and Origins, and have shown that water features are almost completely flattened by clouds, leading to no possible detections by these future observatories. This demonstrates that larger aperture sizes and/or better instrument precision offered by these telescopes cannot compensate for the weakness of the transmission spectroscopy technique to probe tropospheric water.

Furthermore, \citet{fujii:2017} used the ROCKE-3D GCM to study the processes that control the water vapor mixing ratio in the upper atmosphere of synchronously rotating terrestrial planets with ocean-covered surfaces. In particular, they examine the variance in the water vapor mixing ratio for stellar effective temperatures between 3100~K and 5800~K, producing synthetic transmission spectra based on GCM outputs in order to evaluate the detectability of H$_{2}$O signatures. They find that an increase of water vapor mixing ratios in the upper atmosphere leads to an increase of the H$_{2}$O absorption depth by a factor of a few, making the effective altitude of the H$_{2}$O absorption features as high as 15~km when clouds are included. The star they illustrate for this result is GJ 876, which has T$_{\mathrm{eff}} = 3129$~K and a period of 22~days at a flux of 1~S$_\oplus$ (from their Table 2). Its mass is listed as 0.37~M$_\oplus$ and its radius as 0.3761~R$_\oplus$ (from their Table 1). To compare with our results, the closest simulations we can choose are the intermediate cases for the T$_{\mathrm{eff}} = 3300$~K 1400~W/m$^2$ planet (period = 22~days, mass = 0.249~M$_\oplus$, radius = 0.3~R$_\oplus$, flux = 1.029~S$_\oplus$). Using these planets as comparison, both of our studies generally agree that the effective absorption of water vapor is small. For the \SI{2.7}{\micro\metre} feature, \citet{fujii:2017} find a depth of $\sim$2~ppm with clouds, or $\sim$15~km (see their Figure 8), while we report a depth of $\sim$1~ppm or $\sim$2.5~km (the disparity in the km value is reconciled when the different radii of GJ 876 and our T$_{\mathrm{eff}} = 3300$~K are accounted for). Likewise, both studies find that clouds raise the baseline continuum, with the strength of the increase in the baseline proportional to the incident flux, and find differences between cloud properties for fast rotators (non-synchronous as well as ultra-cool dwarfs) versus slower-rotating synchronously-rotating planets around mid-M dwarfs. 

While our study focuses on the climate and detectability aspects of the results from \citet{ravi:2017}, photochemistry also plays a major role. M dwarfs possess strong UV activity, which may effectively photolyze stratospheric H$_{2}$O. A study involving transmission spectroscopy of a terrestrial planet orbiting an M star was conducted by \citet{lincowski:2018}, who used a 1-D photochemical model to simulate a modern Earth-like atmosphere on TRAPPIST-1e. Their results also show that H$_2$O lines are very shallow and their detection with JWST will be extremely challenging. TRAPPIST-1e has an orbital period of 6.1 days and orbits in the HZ of a 2550~K star. Its incident flux is colder than any of those simulated by \citet{ravi:2017} with a flux of 0.6~S$_\oplus$ as opposed to the 0.88~S$_\oplus$ simulated planet (2600~K 1200 W/m$^2$). For a model of TRAPPIST-1e with a 1 bar surface pressure (80\% N$_2$ and 20\% O$_2$), \citet{lincowski:2018} show that the \SI{2.7}{\micro\metre} H$_2$O feature has a depth of $\sim$10~ppm (see their Figure 3). In comparison, our synthesized spectrum for the 2600~K 1200 W/m$^2$ has a \SI{2.7}{\micro\metre} H$_2$O feature depth of $\sim$15~ppm. Given the difference in incident fluxes, a deeper spectral feature than \citet{lincowski:2018} would not be unexpected. However, comparisons are difficult due to the circulation differences between a 1-D model and 3-D GCM as well as the inclusion of photochemistry by \citet{lincowski:2018}. \citet{fauchez:2019} also examined TRAPPIST-1e using a full 3-D GCM (LMD-G, \citet{wordsworth:2011}) combined with the Atmos photochemical model, and examined various atmospheric compositions (modern Earth, Archean Earth and CO$_2$ dominated atmospheres). For their model of modern Earth (the closest to our ocean Earth) and including clouds, they show a feature depth of $\sim$5~ppm for the \SI{6}{\micro\metre} H$_2$O feature (the \SI{2.7}{\micro\metre} feature disappears due to the addition of CO$_2$). Extrapolating from the trends seen in Figure~\ref{fig:delta-all} for the 2600~K star, and given that TRAPPIST-1e receives less insolation from its host star than our lowest-flux scenario, it follows that \citet{fauchez:2019} also report a feature depth that is smaller than ours, as our simulated planets lie in the inner HZ. This difference compared to our simulations could also be due to different convection schemes between ExoCAM and LMD-G leading to less stratospheric water in the latter case \citep{wolf:2015,fauchez:2019b}, or to photochemistry - near the top of atmosphere (TOA), UV radiation massively photodissociates H$_2$O, drastically reducing its concentration. As a result, the H$_2$O absorption is very small, producing only depths of a few ppm. We leave it up to future studies to compare the various GCM models with similar photochemistry schemes to produce a more accurate comparison.

\citet{afrin:2019} also employed a 1-D photochemical model with varied stellar UV to assess whether H$_{2}$O destruction driven by high stellar UV would affect its detectability in transmission spectroscopy, based on the results from \citet{ravi:2017}. They find that as long as the atmosphere is well-mixed up to 1 mbar, UV activity appears to not impact detectability of H$_{2}$O in the transmission spectrum. However, if the H$_{2}$O is not well-mixed in the atmosphere (like in our simulations), H$_{2}$O photodissociation can drastically reduce H$_{2}$O concentration in the upper atmosphere and therefore weaken the H$_{2}$O spectral signature in transit even further \citep{fauchez:2019}. The strength of the spectral features that \citet{afrin:2019} present are similar to that of \citet{ravi:2017} with a peak strength of $\sim$15~ppm for the 3300~K star with clouds considered, compared to $\sim$2~ppm from our results (See Figure 4 in their paper and Figure~\ref{fig:3300} in ours). The reason for this is that \citet{afrin:2019} use a stellar radius of $0.137$ R$_{\odot}$ following \citet{ravi:2017}, which is quite low and inconsistent with the stellar luminosity and temperature of this particular star. This overestimates the strength of the transit spectral features. While this is the dominant effect that reduces the strength of the features, other effects such as the use of gray cloud model, and the non-inclusion of liquid and ice data from the GCM run outputs likely also contribute to the differences.

\subsection{Implications for Observations with Future Observatories}
\label{sec:implications}
The high cloud decks (especially for ice clouds) shown in Figure~\ref{fig:profiles} that drown out water vapor features in the upper atmospheres of these warm, wet planets are responsible for the bleak observability prospects outlined in Section~\ref{sec:results-exptimes}. In terms of the best exposure times, the \SI{2.7}{\micro\metre} feature is the most observable based on the specific atmosphere we have outlined in this work. As stated in Section~\ref{sec:results-exptimes}, when we run exposure time calculations for the three near-infrared features using JWST, the \SI{2.7}{\micro\metre} feature is consistently characterized in the least amount of exposure time. This spectral feature benefits from having both a high depth and a proximity to the peak of the black-body spectrum for M stars. Meanwhile, both the \SI{1.4}{\micro\metre} and \SI{1.8}{\micro\metre} features are comparably smaller in depth, while the \SI{6}{\micro\metre} feature resides too far out in the mid-infrared range to collect enough photons from the illuminating stellar source. However, the atmospheres modeled in this work are limited in compositional complexity, and do not include CO$_2$, which is expected to be very common in the atmospheres of rocky exoplanets. If CO$_2$ was included in the GCM simulations, we expect those spectral lines to overlap or completely cover the \SI{2.7}{\micro\metre} water vapor feature, making it difficult or impossible to detect \citep{fauchez:2019}. For this reason, the remaining features may be more realistic as observing targets. 

However, when we compare each telescope's overall effectiveness in characterizing H$_2$O, instrument and telescope parameters clearly become important factors. Ignoring the noise floor, LUVOIR-A, although observing a smaller feature (\SI{1.8}{\micro\metre}), almost always has shorter exposure times than for JWST at the \SI{2.7}{\micro\metre} feature, due to its enormous mirror (effective diameter of 14.05~m, see Table~\ref{tab:noise}). If the wavelength range and photon-noise sensitivity for LUVOIR-A extended farther into the infrared, we fully expect that the combination of the mirror size and access to the \SI{2.7}{\micro\metre} feature would be the best match for minimizing exposure times for these particular planets. OST is less optimal than JWST because although they both share the same mirror diameter and similar depths for available spectral features, OST only has access to the \SI{6}{\micro\metre} feature, where not enough stellar photons are collected. Thus, for the same stars, JWST always requires a lower exposure time to characterize the \SI{2.7}{\micro\metre} water feature than OST does with the \SI{6}{\micro\metre} feature. For 58\% of our stellar targets, OST has shorter exposure times with its \SI{6}{\micro\metre} feature than LUVOIR-B does with its \SI{1.8}{\micro\metre} feature, implying that in this case the benefit of having a larger feature slightly outweighs the benefit of having a larger telescope. 

Of course, the most realistic aspect of observing these atmospheres that cannot be ignored is the noise floor introduced by the telescope in its environment. Estimates for the JWST noise floor are in the range of $10-20$~ppm for the NIR instruments to 50~ppm for MIRI \citep{greene:2016}. When we introduce a 5~ppm noise floor into the exposure times as detailed in Section~\ref{sec:calc-exptimes}, we find that no ocean Earth around any of our stellar targets is detectable using any telescope. Unlike JWST and OST, LUVOIR-A and -B are unable to detect any feature with a 3~ppm noise floor. With a 5-$\sigma$ detection and a 3~ppm noise floor, a feature depth of at least 15~ppm is required to be detectable. LUVOIR-A and -B only have access to the \SI{1.4}{\micro\metre} and \SI{1.8}{\micro\metre} features, which never reach a transit depth past 10~ppm (see Figure~\ref{fig:all-ppm}). Only 26 stars survive to 3~ppm that can be detectable by JWST; all are very cool and small. This is because, as seen in Figure~\ref{fig:all-ppm}, only the planets orbiting the 2600~K star have relatively large features because of their uniquely low cloud deck and smaller radii. Only eight of these stars that survive a 3~ppm noise floor also have exposure times less than 1000 hours (see Table~\ref{tab:3ppmstars}). However, these exposure times are generally longer than they would be without a noise floor, as the noise floor sometimes shrinks the observable bandwidth of the feature. For example, a notable star that remains detectable at 3~ppm is TRAPPIST-1. Without considering the noise floor, TRAPPIST-1 would have an exposure time of $\sim$254 hours, but with a 3~ppm noise floor, the length of time rises to $\sim$528 hours. Still, this is less than 1000 hours, so if JWST somehow was able to push its noise floor down to 3~ppm, the characterization of TRAPPIST-1 planets would be feasible if they had atmospheres similar to those modeled here $-$ but considering that TRAPPIST-1e only transits once every 6 days with a transit duration of approximately 1 hour, it would take $\sim8.7$ years of constant transit monitoring before a sufficient S/N would be achieved. 

If the community aims for curating better databases for cooler and smaller stars, while pushing for technological advancements in detector capabilities to limit noise floors, then perhaps one day these water vapor features will be detectable through transmission spectroscopy. Should that time come, the exposure times calculated in this work essentially act as a priority list of TESS stars most amenable to characterization. As we discover more terrestrial planets in the Habitable Zones of cool stars, this list can assist the community in making valuable decisions on which rocky planets are worth pursuing through characterization. 

Although the prospects for observing water vapor features on ocean-covered worlds using our next-generation telescopes are not promising, there are still other methods that can characterize the atmospheres of rocky exoplanets. Contrary to transmission spectroscopy for which the starlight is transmitted through the terminator of the planet, reflection and thermal emission spectroscopy would probe the full disk of the planet. Unlike in transmission spectra where clouds clearly hinder spectral characterization, the radiative effect of clouds may give detectable signals in emission and reflection which yield useful information \citep{yang:2014, wolf:2019}. In addition, reflection and thermal emission spectroscopy are not affected by atmospheric refraction, and can probe the lowest level of the atmosphere where most of the water resides. Cloudy areas of planet may be discernible from clear-sky areas, where reflected and emitted radiation can originate from the planet surface and present a different spectral profile. However, future direct imaging missions such as LUVOIR or HabEX would be limited by the inner working angle (IWA) of their instrument, preventing the observation of HZ planets orbiting very close to their host star such as those modeled in this work. Furthermore, reflected and emitted radiation from a planet may change as a function of orbital phase. This may be particularly relevant for tidally-locked planets around M dwarf stars, where permanent day and permanent night sides exist, and can create significant longitudinal gradients in the temperature, cloud, and outgoing radiation fields. Such phase dependent information can be used to characterize terrestrial planet climates \citep{haqq:2018, wolf:2019, kreidberg:2019}. 

\subsection{Future Improvements in GCM Models}
\label{improvements}
To further understand the detectability of terrestrial planets in the Habitable Zones of M stars, we recommend running GCM simulations with more complex atmospheres. Different molecular species, such as CO$_2$, O$_2$ and O$_3$, should be included to not only represent a more realistic planet, but to also see how exposure times are affected by these different compositions. CO$_2$ may cool the stratosphere while absorption by O$_3$ or aerosols may create inversion layers; these effects could potentially make these planetary atmospheres easier to detect. Note that \cite{fauchez:2019} have shown that CO$_2$ could be the best proxy to detect an atmosphere on terrestrial HZ planets due to its strong absorption at \SI{4.3}{\micro\metre}. The detectability of this feature is weakly affected by the presence of clouds because enough CO$_2$ remains above the cloud deck to saturate this feature and still produce a strong, detectable signal. The presence of lighter molecular weight species, such as H$_2$, could also increase the scale height, therefore increasing the depths of the water vapor features. Future 3-D modeling studies should also use prognostic atmospheric chemistry modules in order to self-consistently predict the atmospheric composition while including a rich set of gas species \citep{chen:2018,chen:2019}. In addition, further understanding cloud behavior in the upper atmospheres of incipient runaway planets could potentially lead to lower exposure times. For example, if the ice cloud decks in actuality span less than 1~dec in pressure, it is possible that the feature depths may grow to be equal or even higher than those in the non-runaway states (as seen in the ``zero" model top treatment for Figure~\ref{fig:all-rkm}, although we do not expect this to be realistic).

This study also demonstrates that understanding cloud processes in the upper atmosphere is crucial for understanding transmission spectra of hot and moist exoplanets. In particular, on warm planets the cloud condensation level moves upward in the atmosphere and to lower pressures \citep{leconte:2013, wolf:2015, ravi:2017}. This becomes problematic for many exoplanet GCMs which generally have model tops no higher than $\sim$0.1 mbar, and cloud parameterizations that are benchmarked to tropospheric conditions on Earth. Only very recently have high-model top GCMs begun to be used for studying mesospheric and lower thermospheric processes in Earth-like exoplanet atmospheres \citep{chen:2019}. Still, successfully incorporating higher model tops is only part of the challenge facing the next generation of 3-D climate models. Cloud microphysical parameterizations in GCMs remain rudimentary, but cloud formation, persistence, and radiative properties are highly sensitive to the size, shape, and evolution of cloud particles. Microphysical calculations that self-consistently predict cloud particle size distributions may affect both the overall climate and the resultant transmission spectra. We recommend that future GCM studies of exoplanets work towards increasing the minimum pressure of the model top, and including more sophisticated cloud microphysical schemes.  

\acknowledgments
G. Suissa acknowledges funding from the John Mather Nobel Scholarship. G. Suissa would also like to thank the Goddard Center for Astrobiology's Undergraduate Research Associates in Astrobiology program for supporting her work during the summer of 2018. The authors acknowledge support from GSFC Sellers Exoplanet Environments Collaboration (SEEC), which is funded in part by the NASA Planetary Science Division’s Internal Scientist Funding Model. 

\software{ExoCAM (\url{https://github.com/storyofthewolf/ExoCAM}), ~ PSG \citep{psg:2018}}

\newpage
\appendix

We include a supplementary digital file that lists the exposure times calculated in this work. Our calculations assume an Earth-sized aquaplanet synchronously rotating each of our dataset of 61,513 TESS target stars, for each main water vapor feature using each telescope (JWST, LUVOIR-A, LUVOIR-B, and OST). We also list all of the relevant stellar properties for each target star used in our calculations. Below we include a descriptive table for our attached file. 

\begin{table}[ht]
\centering
\caption{Descriptive table for our attached machine-readable data file.}
\label{tab:datafile}
\begin{tabular}{|c|c|c|c|}
\hline
Column \# & Label & Explanation & Units \\
\hline
1 & ID & TESS Input Catalog identifier & -- \\
2 & TWOMASS & 2MASS Identifier & -- \\ 
3 & V$_{\textrm{mag}}$ & V magnitude & magnitude \\ 
4 & K$_{\textrm{mag}}$ & K magnitude & magnitude \\
5 & TESSflag & TESS Magnitude Flag & -- \\
6 & T$_{\mathrm{eff}}$ & Effective temperature & K \\
7 & rad & Stellar radius & R$_{\odot}$ \\
8 & mass & Stellar mass & M$_{\odot}$ \\ 
9 & lumclass & Luminosity class & -- \\
10 & lum & Stellar luminosity & L$_{\odot}$ \\ 
11 & d & Distance & parsecs \\
12 & properties-flag & Flag whether TIC included T$_{\mathrm{eff}}$, rad, mass, lum & -- \\
13 & jwst2.7-0ppm & Exposure time for JWST, \SI{2.7}{\micro\metre}, no noise floor & hours \\
14 & jwst1.4-0ppm & Exposure time for JWST, \SI{1.4}{\micro\metre}, no noise floor & hours \\
15 & jwst1.8-0ppm & Exposure time for JWST, \SI{1.8}{\micro\metre}, no noise floor & hours \\ 
16 & ost6.0-0ppm & Exposure time for OST, \SI{6.0}{\micro\metre}, no noise floor & hours \\
17 & luva1.4-0ppm & Exposure time for LUVOIR-A, \SI{1.4}{\micro\metre}, no noise floor & hours \\
18 & luva1.8-0ppm & Exposure time for LUVOIR-A, \SI{1.8}{\micro\metre}, no noise floor & hours \\
19 & luvb1.4-0ppm & Exposure time for LUVOIR-B, \SI{1.4}{\micro\metre}, no noise floor & hours \\ 
20 & luvb1.8-0ppm & Exposure time for LUVOIR-B, \SI{1.8}{\micro\metre}, no noise floor & hours \\
21 & jwst2.7-1ppm & Exposure time for JWST, \SI{2.7}{\micro\metre}, 1~ppm noise floor & hours \\
22 & jwst2.7-3ppm & Exposure time for JWST, \SI{2.7}{\micro\metre}, 3~ppm noise floor & hours \\
23 & ost6.0-1ppm & Exposure time for OST, \SI{6.0}{\micro\metre}, 1~ppm noise floor & hours \\
24 & ost6.0-3ppm & Exposure time for OST, \SI{6.0}{\micro\metre}, 3~ppm noise floor & hours \\
25 & luva1.8-1ppm & Exposure time for LUVOIR-A, \SI{1.8}{\micro\metre}, 1~ppm noise floor & hours \\
26 & luvb1.8-1ppm & Exposure time for LUVOIR-B, \SI{1.8}{\micro\metre}, 1~ppm noise floor & hours \\
\hline
\end{tabular}
\end{table}

\bibliography{bib.bib}

\end{document}